\documentclass[transaction]{IEEEtran}
\usepackage{graphicx}
\usepackage{epstopdf}
\usepackage[latin1]{inputenc}
\usepackage{amsmath,amssymb,amscd,latexsym,dsfont}
\usepackage[english]{babel}
\usepackage{tabularx,cite}
\usepackage{graphicx}
\usepackage{algpseudocode}
\usepackage{algorithm}
\usepackage{algorithmicx}
\usepackage{float}
\usepackage{multicol}
\usepackage{psfrag}
\usepackage{comment,psfrag,subfigure,enumerate}
\usepackage{color}
\usepackage{amsthm}
\ifCLASSINFOpdf
\else
\fi
\begin{document}
%
\title{On the Performance of the Relay-ARQ Networks}
\author{\IEEEauthorblockN{Behrooz Makki, Thomas Eriksson, Tommy Svensson, \emph{Senior Member, IEEE}}\\
\thanks{The authors are with Department of Signals and Systems,
Chalmers University of Technology, Gothenburg, Sweden, Email: \{behrooz.makki, thomase, tommy.svensson\}@chalmers.se}
\thanks{This work was supported in part by the Swedish Governmental Agency for Innovation Systems (VINNOVA) within the VINN Excellence Center Chase.}
}

%
\maketitle
\vspace{-0mm}
\begin{abstract}
This paper investigates the performance of relay networks in the presence of hybrid automatic repeat request (ARQ) feedback and adaptive power allocation. The throughput and the outage probability of different hybrid ARQ protocols are studied for independent and spatially-correlated fading channels. The results are obtained for the cases where there is a sum power constraint on the source and the relay or when each of the source and the relay are power-limited individually. With adaptive power allocation, the results demonstrate the efficiency of relay-ARQ techniques in different conditions.
\end{abstract}
%
\IEEEpeerreviewmaketitle
\vspace{-0mm}
\section{Introduction}
\vspace{-0mm}
Relay-assisted communication is one of the promising techniques that have been proposed for the wireless networks. The main idea of a relay network is to improve the data transmission efficiency by implementation of intermediate relay nodes which support the data transmission from a source to a destination. The relay networks have been adopted in the 3GPP long-term evolution advanced (LTE-A) standardization \cite{parkvall} and are expected to be one of the core technologies for the next generation cellular systems.

From another perspective, hybrid automatic repeat request (ARQ) is a well-established approach for wireless networks \cite{revisiontvtrelay1,1327848,revisiontvtrelay2,revisiontvtrelay3,revisiontvtrelay4,throughputdef,noisyARQ,1200407,Arulselvan,6566132,Tcomkhodemun,ARQGlarsson}. The ARQ systems can be viewed as channels with sequential feedback where, utilizing both forward error correction and error detection, the system performance is improved by retransmitting data that has experienced bad channel conditions. Thus, the combination of relay and ARQ improves the performance of wireless systems, because the ARQ makes it possible to use the relay only when it is \emph{needed}.


Due to the fast growth of wireless networks and data-intensive applications in smart phones, green communication via improving the power efficiency is becoming increasingly important for wireless communication. The network data volume is expected to increase by a factor of $2$ every year, associated with $16-20\%$ increase of energy consumption, which contributes about $2\%$ of global $CO_2$ emissions \cite{Gartner}. Hence, from an environmental point of view, minimizing the power consumption is a very important design consideration, and green data transmission schemes must be taken into account for the wireless networks \cite{1321221,5783982,greenref3,greenref4,greenref5,7037211}. Moreover, as most wireless devices operate with limited battery power, it is very important to find ways of maximizing the device lifetime by efficiently utilizing the limited power. These are the main motivations for this paper, in which we analyze the power-limited performance of the relay-ARQ setups.

The basic principles of different ARQ protocols are derived in \cite{revisiontvtrelay1,1327848,revisiontvtrelay2,revisiontvtrelay3,revisiontvtrelay4,throughputdef,noisyARQ}. Power allocation in ARQ-based single-user (without relay) networks is addressed by, e.g., \cite{1200407,Arulselvan,6566132,Tcomkhodemun,ARQGlarsson}. Also, \cite{a04133862,a4450823,a5683506,05502432,relay12,relay14,relay15,relay16} study the problem in relay networks. There are a number of papers dealing with energy efficiency and power allocation in relay-ARQ setups. These works can be divided into two categories, as stated in the following.

In \cite{a04786513,a06133849,a06051530,a06184258,a05956564,a05992825,a06213570,ARQcollaborative2006,lanati2,ARQrelaynarasimhan,ARQcooperativeglobcom,6477555,6512535}, the source and the relay use, e.g., space-time codes (STCs) to make a distributed cooperative antenna and retransmit the data simultaneously in rounds when the relay is active; With an outage probability constraint, \cite{a04786513,a06133849} (resp. \cite{a06051530}) study the energy efficiency (resp. long-term average transmission rate) of STC-based relay-ARQ systems. The energy and spectrum efficiency of the basic and hybrid relay-ARQ networks are verified in \cite{a06184258,a05956564,a05992825} as well. Also, \cite{a06213570} designs a multi-relay-ARQ network using Alamouti codes. Assuming the source and the relay to be close, \cite{ARQcollaborative2006,lanati2} investigate the throughput of relay networks using different ARQ protocols. Optimizing the delay-limited throughput and deriving a closed-form expression for the average power of the source are addressed by \cite{ARQrelaynarasimhan} and \cite{ARQcooperativeglobcom}, respectively. Considering the incremental redundancy (INR) protocol, \cite{6477555} studies the performance of the relay-ARQ setups in fast-fading conditions. Finally, the results of \cite{6477555} are extended in \cite{6512535}, where the system performance is compared with cases having only one of the source or the relay active in the retransmissions. Implementation of STCs in these works is based on the assumption that there is perfect synchronization between the source and the relay.

In \cite{powerarq2007,a06399140,a06214030,a06399491,a06409500,milicomARQrelay,elsevierARQ2007,a06247450,a06188994,a06362532,a06189805}, only one terminal (either the source or the relay) is active in the retransmission rounds, as opposed to \cite{a04786513,a06133849,a06051530,a06184258,a05956564,a05992825,a06213570,ARQcollaborative2006,lanati2,ARQcooperativeglobcom,ARQrelaynarasimhan}. For instance, \cite{powerarq2007} studies the outage-limited energy minimization in single-user and relay-ARQ networks. Opportunistic relaying, rate adaptation and analyzing the energy-delay tradeoff curve are considered by \cite{a06399140}, \cite{a06214030} and \cite{a06399491,a06409500}, respectively, where the direct source-destination link is ignored. Also, the throughput, the packet error rate and the effective capacity of different ARQ-assisted relay networks are studied in \cite{milicomARQrelay,elsevierARQ2007,a06247450}, respectively.
Power scaling in MIMO and cognitive radio relay-ARQ networks is addressed in \cite{a06188994} and \cite{a06362532}, respectively. Finally, \cite{a06189805} studies a relay-ARQ network using superposition coding. References \cite{6477555,6512535,a06362532,a05956564, a05992825,a06189805,a04786513,a06133849,a06399491,a06409500,a06051530,a06184258,a06213570,ARQcollaborative2006,a06188994,lanati2,ARQrelaynarasimhan,a06399140,a06214030,milicomARQrelay,elsevierARQ2007,a06247450} are based on the assumption that there is a fixed transmission power for the source and the relay. Meanwhile \cite{a06051530, 05502432,a06399491,a06409500} optimize the power allocation between the source and the relay under a sum power constraint, while they use the same powers in all retransmissions. Also, \cite{powerarq2007} investigates the power allocation between the retransmissions for basic ARQ schemes and \cite{ARQcooperativeglobcom} studies the average power of the source with repetition time diversity (RTD) ARQ and a fixed power for the relay.

In theoretical investigations, the communication links between the source, the relay and the destination are normally assumed to be independent \cite{powerarq2007,a06362532,a05956564,6512535,6477555, a05992825,a06189805,a04786513,a06133849,a06399491,a06409500,a06051530,a06184258,a06213570,ARQcollaborative2006,a06188994,lanati2,ARQcooperativeglobcom,ARQrelaynarasimhan,a06399140,a06214030,milicomARQrelay,elsevierARQ2007,a06247450,a06152804}. This is an appropriate model for many practical scenarios \cite{powerarq2007,a06362532,a05956564,6477555,6512535, a05992825,a06189805,a04786513,a06133849,a06399491,a06409500,a06051530,a06184258,a06213570,ARQcollaborative2006,a06188994,lanati2,ARQcooperativeglobcom,ARQrelaynarasimhan,a06399140,a06214030,milicomARQrelay,elsevierARQ2007,a06247450,a06152804} and makes it possible to analyze the system performance analytically. However, the independent fading channel is not always a realistic model. For instance, the relay is normally located close to the destination in \emph{moving-relay} systems \cite{6240247,6379135}. As a result, there might be considerable correlation between the source-relay and the source-destination fading coefficients. Also, e.g., \cite{a06189805} demonstrates the cases where the source is connected to the destination through a relay which is close to the source.
In this case, the source-destination and the relay-destination links may be spatially-correlated. For these reasons, it is interesting to extend the independent fading model to the case where there is spatial correlation between the channels.

In this paper, we study
the throughput and the outage probability of the relay-ARQ networks in cases where there is either a long-run sum power constraint on the source and the relay or when each of the source and the relay are power-limited individually. Adaptive power allocation between the retransmissions is used to improve the system performance. We derive closed-form expressions for the average power, the throughput and the outage probability of different relay-ARQ protocols in the cases with independent or spatially-correlated fading channels. Moreover, we investigate the effect of fading temporal variations on the data transmission efficiency of the relay-ARQ systems.

As opposed to \cite{a04786513,a06133849,6477555,a06051530,a06184258,a05956564,a05992825,a06213570,ARQcollaborative2006,lanati2,ARQcooperativeglobcom,ARQrelaynarasimhan}, we study the scenario where only one of the source or the relay is active in each ARQ-based retransmission round. Also, the problem setup of the paper is different from the ones in \cite{a04786513,a06133849,a06051530,6477555,6512535,a06184258,a05956564,a05992825,a06213570,ARQcollaborative2006,lanati2,ARQcooperativeglobcom,ARQrelaynarasimhan,powerarq2007,a06399140,a06214030,a06399491,a06409500,milicomARQrelay,elsevierARQ2007,a06247450,a06188994,a06362532,a06189805} because 1) we consider adaptive power allocation between retransmissions of hybrid ARQ protocols, 2) the results are obtained with different sum and individual power constraints on the source and the relay and 3) we investigate the system performance in both independent and spatially-correlated fading conditions, with noisy/noise-free feedback signals. Finally, our discussions on the users' message decoding probabilities (Theorems 1-3) have not been presented before.

The results show that there is a structural procedure to study different performance metrics of relay-ARQ networks experiencing different fading models.
Optimal power allocation is shown to be very useful in terms of outage probability, throughput and coverage region of the relay-ARQ network, when there is a sum power constraint on the source and the relay. With individual power constraints on the source and the relay, however, optimal power allocation increases the throughput (resp. reduces the outage probability) only at low (resp. high) signal-to-noise ratios (SNRs). Compared to the fixed-length coding scheme, the throughput of the relay-ARQ network increases when variable-length coding is utilized. With the practical range of spatial correlations, the performance of the relay-ARQ network is not sensitive to the spatial correlation. However, the data transmission efficiency of the network is reduced at highly-correlated conditions.
 \vspace{-3mm}
\section{System model}
We consider a relay-assisted communication setup consisting of a source, a relay and a destination. The channel coefficients in the source-relay, the source-destination and the relay-destination links are denoted by $h^\text{sr},h^\text{sd}$ and $h^\text{rd}$, respectively. Also, we define $g^\text{sr}\doteq |h^\text{sr}|^2,$ $g^\text{sd}\doteq |h^\text{sd}|^2$ and $g^\text{rd}\doteq |h^\text{rd}|^2$ which are referred to as the channel gains in the following. A maximum number of $M$ ARQ-based retransmission rounds is considered, i.e., the data is (re)transmitted a maximum of $M+1$ times. Moreover, we define a packet as the transmission of a codeword along with all its possible retransmission rounds. In each packet, $Q$ information nats are sent to the destination and the length of the subcodeword used in the $m$-th round of the ARQ is denoted by $l_m$. Thus, the equivalent data rate, i.e., the code rate of the ARQ, at the end of the $m$-th round is given by $R_{(m)}=\frac{Q}{\sum_{n=1}^m{l_n}}.$

We study the system performance for two different block-fading conditions:
\begin{itemize}
  \item Quasi-static. In this model, the channel coefficients are assumed to remain fixed within a packet period, and then change to other values based on their probability density functions (pdf).
  \item Fast-fading. Here, the channel coefficients are supposed to change in each retransmission round.
\end{itemize}


The quasi-static model, studied in Subsections IV.A-B, represents the scenarios with slow-moving or stationary users, e.g., \cite{Tcomkhodemun,6566132,1661837}. On the other hand, the fast-fading, studied in Subsection IV.C, is an appropriate model for the high speed users and frequency-hopping setups where the channel quality changes in the retransmissions independently, e.g., \cite{6512535,1661837,ARQGlarsson}.


In each link, the channel coefficient is assumed to be known by the receiver, which is an acceptable assumption in block-fading channels \cite{1200407,Arulselvan,6566132,Tcomkhodemun,ARQGlarsson}. However, there is no instantaneous channel state information available at the transmitters except the ARQ feedback bits.
The ARQ feedback signals are initially assumed to be received error-free, but we later investigate the effect of erroneous feedback bits as well (Section IV).

\emph{Relay-ARQ model:} The considered relay-ARQ protocol works as follows. In each packet period, the data transmission starts from the source. If the data is decoded by the destination, an acknowledgement (ACK) is fed back by the destination to the source and the relay, and the retransmissions stop. Otherwise, the destination transmits a negative-acknowledgment (NACK). Only one terminal (either the source or the relay) is active in each retransmission round; the relay becomes active and the source turns off, as soon as the data is decoded by the relay. That is, if the relay successfully decodes the message, it sends an ACK to the source and starts retransmission until the destination decodes the data or the maximum number of retransmissions is reached. In other words, with error-free feedback bits, the following cases may occur during a packet transmission period: 1) receiving an ACK from the relay and a NACK from the destination, the source turns off and the relay starts retransmission. 2) With NACKs from the relay and the destination, the data is retransmitted by the source. 3) Receiving an ACK from the destination, the source ignores the ARQ feedback of the relay and the {re}transmissions stop (Performance analysis in the cases with noisy feedback bits is studied in Section IV.D.).

The motivations for considering the proposed data transmission model are as follows. Letting the relay retransmit instead of the source when the source-destination link experiences bad condition makes it possible to exploit the potential diversity gain through the relay channel. Also, in practice, the relay is located such that the relay-destination link experiences better \emph{average} characteristics than the source-destination link. Therefore, it is more beneficial to use the power resources for the relay, instead of dividing the power between the source and the relay, if the relay decodes the message. Finally, as seen in the following, for Rayleigh-fading conditions the proposed scheme outperforms the state-of-the-art approaches, in terms of outage probability/throughput.
 \vspace{-0mm}
\section{Problem formulation}
 \vspace{-0mm}
In this paper, we study the problem of
 \vspace{-0mm}
\begin{align}\label{eq:optproblem}
\begin{array}{l}
 \mathop {\max }\limits_{\forall R_{(m)},P_m^\text{s},P_m^\text{r}} \,\, \Omega,\,\,\,\, \Omega=\{\eta,\,-\Pr(\text{Outage})\} \\
 \,\,\,\text{subject}\,\text{to}\,\,\,\,  \Delta, \,\, \,\, \Delta=\{\Phi^\text{total}\le\phi^\text{total},\,\,(\Phi^\text{s}\le\phi^\text{s}) \& (\Phi^\text{r}\le\phi^\text{r})\}.  \\
 \end{array}
\end{align}
In words, we investigate the long-term throughput $\eta$ and the outage probability $\Pr(\text{Outage})$ as the evaluation yardsticks. The optimization parameters are the equivalent data transmission rates $R_{(m)}$ as well as $P_m^\text{s}$ and $P_m^\text{r},m=1,\ldots,M+1,$
which denote the source and the relay power used in the $m$-th retransmission round, respectively (because the noise variances are set to 1, $P_m^\text{s}$ and $P_m^\text{r}$, in dB $10\log_{10}(P_m^\text{s})$ and $10\log_{10}(P_m^\text{r})$, represent the transmission SNR as well). Finally, the throughput and the outage probability are optimized under two different power-limited scenarios:
\begin{itemize}
  \item \emph{Scenario 1.} The total power for data transmission in the relay-ARQ setup is limited, which is represented by $\Phi^\text{total}\le\phi^\text{total}$. Here, $\Phi^\text{total}$ is the total power in the source and the relay, averaged over many packet transmissions, and $\phi^\text{total}$ denotes the total power constraint. This scenario is of interest in the green communication concept, where the goal is to minimize the total average power required for data transmission \cite{1321221,5783982,greenref3,greenref4,greenref5,7037211}, and also for electricity-bill minimization.
  \item \emph{Scenario 2.} There are individual power constraints on the source and the relay, which is represented by $(\Phi^\text{s}\le\phi^\text{s}) \& (\Phi^\text{r}\le\phi^\text{r})$ in (\ref{eq:optproblem}). Here, $\Phi^\text{s}$ and $\Phi^\text{r}$ are the average power in the source and the relay, respectively, and $\phi^\text{s}$ and $\phi^\text{r}$ denote their corresponding thresholds. This scenario models the case where the source and the relay are battery-limited \cite{6566132,1200407,Arulselvan,ARQGlarsson,Tcomkhodemun}.
\end{itemize}
\begin{figure}\label{fig:NPA}
\vspace{-0mm}
\centering
  \includegraphics[width=.98\columnwidth]{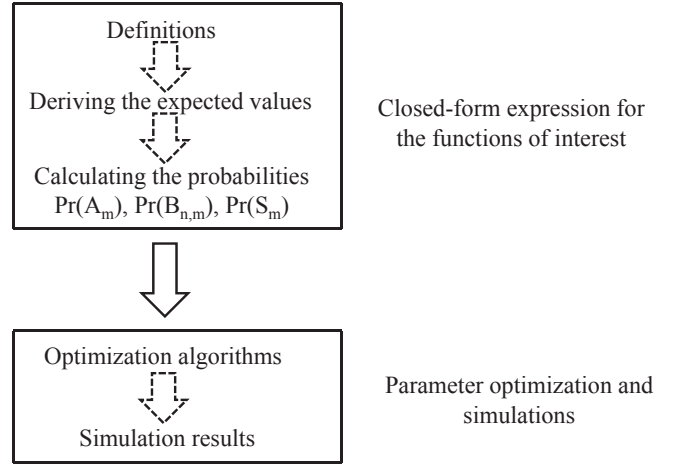}\\\vspace{-0mm}
\caption{An overview of the paper. First, we obtain closed-form expressions for the functions involved in (\ref{eq:optproblem}). Then, iterative optimization algorithms are applied to optimize the parameters based on the closed-form expressions. }\label{figure111}
\vspace{-0mm}
\end{figure}
To study (\ref{eq:optproblem}), the following procedure is considered (please see Fig. 1 as well). First, we derive closed-form expressions for the functions $\eta$, $\Pr(\text{Outage})$, $\Phi^\text{total}$, $\Phi^\text{s}$ and $\Phi^\text{r}$ which are involved in (\ref{eq:optproblem}).
Then, since (\ref{eq:optproblem}) is a nonconvex problem, iterative optimization algorithms are used to optimize the parameters based on the closed-form expressions.

In three steps we obtain the closed-form expressions of $\eta$, $\Pr(\text{Outage})$, $\Phi^\text{total}$, $\Phi^\text{s}$ and $\Phi^\text{r}$. The first step is to define the metrics and the constraints as functions of a few expected values. Then, in the second step, we derive the expectations as functions of predefined probability terms. The last step is to represent the considered probabilities in terms of $R_{(m)},\, P_m^\text{s},\,P_m^\text{r},\,m=1,\ldots,M+1,$ i.e., the optimization parameters of (\ref{eq:optproblem}). Interestingly, the two first steps are independent of the considered ARQ protocol and the fading channel model. Thus, they are explained in the two following subsections. The third step, however, depends on the characteristics of the ARQ schemes and the fading channel model. For this reason, we specify the results for different ARQ protocols and fading channel models in Sections IV and V.
\vspace{-0mm}
\subsection{Step 1: Definitions}

The outage probability is defined as the probability of the event that the data can not be decoded by the destination when the data (re)transmission is stopped. Also, the throughput (in nats per channel use (npcu)) is given by \cite{revisiontvtrelay2,6566132,Tcomkhodemun,throughputdef}
\vspace{-0mm}
\begin{align}\label{eq:throughputdef}
\eta=\mathop {\lim }\limits_{K \to \infty } \frac{{{\sum_{k=1}^{K}{Q_k}}}}{{{\sum_{k=1}^{K}{t_k^\text{total}}}}}=\mathop {\lim }\limits_{K \to \infty } \frac{{\frac{1}{K}{\sum_{k=1}^{K}{Q_k}}}}{{\frac{1}{K}{\sum_{k=1}^{K}{t_k^\text{total}}}}}\mathop  = \limits^{(a)}\frac{E\{\mathcal{Q}\}}{E\{\mathcal{T}^\text{total}\}}.
\end{align}
Here, $Q_k$ is the number of information nats successfully decoded by the destination in the $k$-th packet transmission. Also, $t_k^\text{total}$ denotes the total number of channel  uses in the $k$-th packet transmission, i.e., $t_k^\text{total}={\sum_{n=1}^m{l_n}}$ if the message retransmission of the $k$-th packet continues for $m$ rounds (see (\ref{eq:expectedchanneluses})). Note that in each packet part of the data may be (re)transmitted by the source or the relay and $t_k^\text{total}=t_k^\text{s}+t_k^\text{r}$ where $t_k^\text{s}$ and $t_k^\text{r}$ are the source and the relay activation periods in the $k$-th packet transmission, respectively. In general, $Q_k$ and $t_k^\text{total}$ are random values which follow the random variables $\mathcal{Q}$ and $\mathcal{T}^\text{total}$, respectively, as functions of the channel realizations. Also, $(a)$ in (\ref{eq:throughputdef}) is based on the law of large numbers, with $E\{.\}$ representing the expectation operator, and the fact that the limits, e.g., $\mathop {\lim }\limits_{K \to \infty } \frac{1}{K}\sum_{k=1}^{K}{Q_k}$, $\mathop {\lim }\limits_{K \to \infty } \frac{1}{K}\sum_{k=1}^{K}{t_k^\text{total}}$, exist \cite{6566132,Tcomkhodemun,throughputdef}.

With the same arguments, the average power terms $\Phi^\text{total}$, $\Phi^\text{s}$ and $\Phi^\text{r}$ are obtained by
\vspace{-0mm}
\begin{align}\label{eq:powersource}
\Phi^\text{s}=\mathop {\lim }\limits_{K \to \infty } \frac{{{\sum_{k=1}^{K}{\xi_k^\text{s}}}}}{{{\sum_{k=1}^{K}{t_k^\text{s}}}}}=\frac{E\{\Xi^\text{s}\}}{E\{\mathcal{T}^\text{s}\}},
\end{align}
\vspace{-0mm}
\begin{align}\label{eq:powerrelay}
\Phi^\text{r}=\mathop {\lim }\limits_{K \to \infty } \frac{{{\sum_{k=1}^{K}{\xi_k^\text{r}}}}}{{{\sum_{k=1}^{K}{t_k^\text{r}}}}}=\frac{E\{\Xi^\text{r}\}}{E\{\mathcal{T}^\text{r}\}},
\end{align}
\vspace{-0mm}
\begin{align}\label{eq:powertotal}
\Phi^\text{total}&=\mathop {\lim }\limits_{K \to \infty } \frac{{{\sum_{k=1}^{K}{\xi_k^\text{total}}}}}{{{\sum_{k=1}^{K}{t_k^\text{total}}}}}=\mathop {\lim }\limits_{K \to \infty } \frac{{{\sum_{k=1}^{K}{\xi_k^\text{s}}}+{\sum_{k=1}^{K}{\xi_k^\text{r}}}}}{{{\sum_{k=1}^{K}{t_k^\text{total}}}}}\nonumber\\&=\frac{E\{\Xi^\text{s}\}+E\{\Xi^\text{r}\}}{E\{\mathcal{T}^\text{total}\}}.
\end{align}
Here, $\xi_k^\text{s},\,\xi_k^\text{r}$ and $\xi_k^\text{total}$ are the source, the relay and the total transmission energy in the $k$-th packet transmission, respectively, with $\xi_k^\text{total}=\xi_k^\text{s}+\xi_k^\text{r}$. Also, $\Xi^\text{s},\, \Xi^\text{r},\, \mathcal{T}^\text{s}$ and $\mathcal{T}^\text{r}$ denote the random variables corresponding to $\xi_k^\text{s},\, \xi_k^\text{r},\, t_k^\text{s}$ and $t_k^\text{r}$, respectively.
Note that the metrics and constraints are functions of a few expected values.
\vspace{-0mm}
\subsection{Step 2: Deriving the Expected Values}
Let us define the following events:
\begin{itemize}
  \item $A_m$ is the event that the data is successfully decoded by the destination in the $m$-th (re)transmission round \emph{while it was not decodable before}. In this case, the codeword may have been sent by the source or relay.
  \item $B_{n,m}$ represents the event that the relay is active in rounds $n+1,\ldots,m$ with $n=1,\ldots,M,n<m.$ In this case, the source message has been decoded by the relay in the $n$-th round and, consequently, the source turns off in the successive retransmissions. The relay data retransmission is stopped in the $m$-th round if the destination can decode the data or the maximum number of retransmissions is reached.
  \item $S_m$ is the event that the source stops data retransmission in the $m$-th round. In this case, either the maximum number of retransmissions is reached or the data has been decoded by the relay or the destination.
\end{itemize}
The defined events are used to express (\ref{eq:throughputdef})-(\ref{eq:powertotal}) as functions of $R_{(m)},$ $P_m^\text{s}$ and $P_m^\text{r},\,m=1,\ldots,M+1.$ The details are explained as follows.

According to the definitions, the outage probability is found as
\vspace{-2mm}
\begin{align}\label{eq:outageprob}
\Pr(\text{Outage})=1-\sum_{m=1}^{M+1}{\Pr(A_m)}.
\end{align}
If the data is decoded by the destination at any (re)transmission round, all $Q$ information nats of the packet are received. Hence, the expected number of received information nats in each packet is
\vspace{-2mm}
\begin{align}\label{eq:expectedQnats}
{{E}}\{\mathcal{Q}\}=Q\left(1-\Pr(\text{Outage})\right)=Q\sum_{m=1}^{M+1}{\Pr(A_m)}.
\end{align}
If the data is decoded at the end of the $m$-th round, the total number of channel uses is $l_{(m)}=\sum_{i=1}^{m}{l_i}$.
Also, the total number of channel uses is $l_{(M+1)}=\sum_{i=1}^{M+1}{l_i}$ if an outage occurs, where all possible retransmission rounds are used. Thus, the expected number of total channel uses, i.e., ${\text{E}}\{\mathcal{T}^\text{total}\}$ in (\ref{eq:throughputdef}) and (\ref{eq:powertotal}), is obtained by
\begin{align}\label{eq:expectedchanneluses}
{{E}}\{\mathcal{T}^\text{total}\}  = \sum_{m = 1}^{M + 1} {\left(\sum_{i = 1}^m {{l_i}} \right)\Pr (A_{m})}  + (\sum_{i = 1}^{M + 1} {{l_i}} )\Pr (\text{Outage}).
\end{align}
From (\ref{eq:expectedQnats})-(\ref{eq:expectedchanneluses}) and $R_{(m)}=\frac{Q}{\sum_{i=1}^{m}{l_i}}$, the throughput (\ref{eq:throughputdef}) is found as
\begin{align}\label{eq:throughputasli}
\eta=\frac{\sum_{m=1}^{M+1}{\Pr(A_m)}}{\sum_{m = 1}^{M + 1} {\frac{\Pr (A_{m})}{R_{(m)}}}  + \frac{1-\sum_{m=1}^{M+1}{\Pr(A_m)}}{R_{(M+1)}}}.
\end{align}

If the source stops data (re)transmission at the end of the $m$-th round, the total energy consumed by the source is $\xi_{(m)}^\text{s}=\sum_{i=1}^{m}{P_i^\text{s}l_i}$. Therefore, the source consumed energy is a random variable given by
\begin{align}\label{eq:sourceenergyrv}
\Xi^\text{s}=\sum_{i=1}^{m}{P_i^\text{s}l_i},\,\,\,\, \text{if  } S_m,\,m=1,\ldots,M+1,
\end{align}
and, using $l_i=Q(\frac{1}{R_{(i)}}-\frac{1}{R_{(i-1)}}),$ we have
\begin{align}\label{eq:sourceexpectedenergy}
E\{\Xi^\text{s}\}&=\sum_{m=1}^{M+1}{\left(\left(\sum_{i=1}^{m}{P_i^\text{s}l_i}\right)\Pr(S_m)\right)}\nonumber\\&=Q\sum_{m=1}^{M+1}{\left(\left(\sum_{i=1}^{m}{P_i^\text{s}\left(\frac{1}{R_{(i)}}-\frac{1}{R_{(i-1)}}\right)}\right)\Pr(S_m)\right)}.
\end{align}
With the same arguments, the expected activation period of the source, i.e., $E\{\mathcal{T}^\text{s}\}$ in (\ref{eq:powersource}), is found as
\begin{align}\label{eq:sourceexpectedchanneluse}
E\{\mathcal{T}^\text{s}\}=\sum_{m=1}^{M+1}{\left(\left(\sum_{i=1}^{m}{l_i}\right)\Pr(S_m)\right)}=Q\sum_{m=1}^{M+1}{\frac{\Pr(S_m)}{R_{(m)}}}
\end{align}
which, along with (\ref{eq:sourceexpectedenergy}), leads to
\begin{align}\label{eq:sourcepowerasli}
\Phi^\text{s}=\frac{\sum_{m=1}^{M+1}{\left(\left(\sum_{i=1}^{m}{P_i^\text{s}\left(\frac{1}{R_{(i)}}-\frac{1}{R_{(i-1)}}\right)}\right)\Pr(S_m)\right)}}{\sum_{m=1}^{M+1}{\frac{\Pr(S_m)}{R_{(m)}}}}.
\end{align}
Given that the data is retransmitted by the relay in the $n+1,\ldots,m$ rounds, its consumed energy is $\xi_{(n+1,m)}^\text{r}=\sum_{i=n+1}^{m}{P_i^\text{r}l_i}$ which is consumed during $\sum_{i=n+1}^{m}{l_i}$ channel uses. Thus, we can use the definition of $B_{n,m},$ the same procedure as in (\ref{eq:sourceenergyrv})-(\ref{eq:sourcepowerasli}) and $\sum_{j=n+1}^{m}{l_j}=Q(\frac{1}{R_{{(m)}}}-\frac{1}{R_{{(n)}}})$ to write
\begin{align}\label{eq:relayexpectedenergy}
&E\{\Xi^\text{r}\}={\sum_{\forall n<m,m\le M+1}{\xi_{(n+1,m)}^\text{r}\Pr(B_{n,m})}} \nonumber\\&=Q{\sum_{\forall n<m,m\le M+1}{\left(\left(\sum_{i=n+1}^{m}{P_i^\text{r}\left(\frac{1}{R_{(i)}}-\frac{1}{R_{(i-1)}}\right)}\right)\Pr(B_{n,m})\right)}},
\end{align}
\begin{align}\label{eq:relayexpectedchanneluses}
E\{\mathcal{T}^\text{r}\}&={\sum_{\forall n<m,m\le M+1}{(\sum_{i=n+1}^{m}{l_i})\Pr(B_{n,m})}}\nonumber\\&=Q{\sum_{\forall n<m,m\le M+1}{\left(\Pr(B_{n,m})(\frac{1}{R_{{(m)}}}-\frac{1}{R_{{(n)}}})\right)}},
\end{align}
\begin{align}\label{eq:relaypowerasli}
&\Phi^\text{r}=\frac{E\{\Xi^\text{r}\}}{E\{\mathcal{T}^\text{r}\}}=\nonumber\\&\frac{\sum_{\forall n<m,m\le M+1}{\left(\left(\sum_{i=n+1}^{m}{P_i^\text{r}\left(\frac{1}{R_{(i)}}-\frac{1}{R_{(i-1)}}\right)}\right)\Pr(B_{n,m})\right)}}{\sum_{\forall n<m,m\le M+1}{\left(\Pr(B_{n,m})(\frac{1}{R_{{(m)}}}-\frac{1}{R_{{(n)}}})\right)}}.
\end{align}
Note that the summations in (\ref{eq:relayexpectedenergy})-(\ref{eq:relaypowerasli}) are on all possible activation conditions of the relay. Finally, from (\ref{eq:powertotal}), (\ref{eq:expectedchanneluses}), (\ref{eq:sourceexpectedenergy}), (\ref{eq:relayexpectedenergy}), the total transmission power $\Phi^\text{total}=\frac{E\{\Xi^\text{s}\}+E\{\Xi^\text{r}\}}{E\{\mathcal{T}^\text{total}\}}$ is obtained by
\begin{align}\label{eq:totalpower}
&\Phi^\text{total}=\frac{\varpi}{\sum_{m = 1}^{M + 1} {\frac{\Pr (A_{m})}{R_{(m)}}}  + \frac{1-\sum_{m=1}^{M+1}{\Pr(A_m)}}{R_{(M+1)}}},\nonumber\\&
\varpi\doteq\sum_{m=1}^{M+1}{\left(\sum_{i=1}^{m}{P_i^\text{s}\left(\frac{1}{R_{(i)}}-\frac{1}{R_{(i-1)}}\right)}\right)\Pr(S_m)}\nonumber\\&+{\sum_{\forall n<m,m\le M+1}{\left(\left(\sum_{i=n+1}^{m}{P_i^\text{r}\left(\frac{1}{R_{(i)}}-\frac{1}{R_{(i-1)}}\right)}\right)\Pr(B_{n,m})\right)}}.
\end{align}

From (\ref{eq:outageprob})-(\ref{eq:totalpower}), it follows that the only difference between different ARQ protocols is in the probability terms $\Pr(A_m),\,\Pr(S_m)$ and $\Pr(B_{n,m})$. Also, to derive closed-form expressions for the outage probability, the throughput and the average power terms, the final step is to represent the probabilities $\Pr(A_m),$ $\Pr(S_m)$ and $\Pr(B_{n,m})$ as functions of $R_{(m)},\, P_m^\text{s},\,P_m^\text{r},\,m=1,\ldots,M+1,$ i.e., the optimization parameters of (\ref{eq:optproblem}). Sections IV and V are devoted to obtain the probability terms for different ARQ protocols and fading channel models.
\vspace{-0mm}
\section{Performance analysis in spatially-independent Rayleigh-fading conditions}

For the spatially-independent Rayleigh-fading channels the fading coefficients follow $h^\text{sr}\sim \mathcal{CN}(0,\frac{1}{\lambda^\text{sr}}),\,h^\text{sd}\sim \mathcal{CN}(0,\frac{1}{\lambda^\text{sd}})$ and $h^\text{rd}\sim \mathcal{CN}(0,\frac{1}{\lambda^\text{rd}}).$ Thus, the pdf of the channel gains are given by $f_{g^\text{sr}}(x)=\lambda^\text{sr}e^{-\lambda^\text{sr}x},$ $f_{g^\text{sd}}(x)=\lambda^\text{sd}e^{-\lambda^\text{sd}x}$ and $f_{g^\text{rd}}(x)=\lambda^\text{rd}e^{-\lambda^\text{rd}x}.$ Here, $\lambda^\text{sr},\,\lambda^\text{rd}$ and $\lambda^\text{sd}$ are the fading parameters determined by the path loss and shadowing between the corresponding terminals. Performance analysis of the relay-ARQ setup in the presence of RTD and INR protocols is as follows.
\vspace{-0mm}
\subsection{RTD Protocol in Quasi-Static Conditions}
Using the RTD protocol, the same codeword is (re)transmitted in each round and the receiver performs maximum ratio combining (MRC) of the received signals.
Thus, the equivalent data rate at the end of the $m$-th round is $R_{(m)}=\frac{Q}{ml}=\frac{R}{m}$ with $R$ and $l$ representing the initial data rate and the length of the codeword, respectively. Moreover, with MRC, the received SNR of, e.g., the relay at the end of the $m$-th retransmission round increases to $g^\text{sr}\sum_{i=1}^{m}{P_i^\text{s}}.$ Thus, the data is decoded by the relay at the end of the $m$-th round (and not before) if  $\log(1+g^\text{sr}\sum_{i=1}^{m-1}{P_i^\text{s}})<R\le \log(1+g^\text{sr}\sum_{i=1}^{m}{P_i^\text{s}}),$ which is based on the fact that with SNR $x$ the maximum decodable rate is $\frac{1}{m}\log(1+x),$ if a codeword is repeated $m$ times \cite{Tcomkhodemun,throughputdef}. In this way, $\Pr(S_m)$, i.e., the probability that the source stops retransmission at round $m$, is found as
\begin{align}\label{eq:RTDSm}
&\Pr(S_m) = \left\{ \begin{array}{l}
 {\alpha_m+\beta_m}\,\,\,\,\,\,\,\,\text{if}\,\,m=1,\ldots,M, \\
 \gamma_{M}\,\,\,\,\,\,\,\,\,\,\,\,\,\,\,\,\,\,\,\,\,\,\text{if}\,\,m=M+1,\nonumber \\
 \end{array} \right.
\\
&\alpha_m=\Pr(\log(1+g^\text{sr}\sum_{i=1}^{m-1}{P_i^\text{s}})<R\nonumber\\&\,\,\,\,\,\,\,\,\,\,\,\,\,\,\,\,\,\,\,\,\,\,\le \log(1+g^\text{sr}\sum_{i=1}^{m}{P_i^\text{s}})\,\cap\,\log(1+g^\text{sd}\sum_{i=1}^{m}{P_i^\text{s}})<R),\nonumber
\\
&\beta_m=\Pr(\log(1+g^\text{sd}\sum_{i=1}^{m-1}{P_i^\text{s}})<R\nonumber\\&\,\,\,\,\,\,\,\,\,\,\,\,\,\,\,\,\,\,\,\,\,\,\le \log(1+g^\text{sd}\sum_{i=1}^{m}{P_i^\text{s}})\,\cap\,\log(1+g^\text{sr}\sum_{i=1}^{m-1}{P_i^\text{s}})<R),\nonumber
\\
&\gamma_M={\Pr(\log(1+g^\text{sd}\sum_{i=1}^{M}{P_i^\text{s}})<R\,\cap\,\log(1+g^\text{sr}\sum_{i=1}^{M}{P_i^\text{s}})<R)}.
\end{align}
Here, $\alpha_m$ is the probability that the relay decodes the data at round $m$, before the destination. Moreover, $\beta_m$ denotes the probability that the destination decodes the data at round $m$ while the relay had not decoded the message up to the end of the $(m-1)$-th round (the message may be decodable by the relay at the $m$-th round). Also, the source retransmits the codeword $M+1$ times, if none of the relay and the destination have decoded the data until the $M$-th round, leading to $\gamma_M$ in (\ref{eq:RTDSm}). Note that $\sum_{m=1}^{M+1}{\Pr(S_m)}=1,$ as a maximum of $M+1$ (re)transmissions is considered. For independent Rayleigh-fading channels, (\ref{eq:RTDSm}) is rephrased as
\begin{align}
&\alpha_m=\left(F_{g^\text{sr}}(\frac{e^R-1}{\sum_{i=1}^{m-1}{P_i^\text{s}}})-F_{g^\text{sr}}(\frac{e^R-1}{\sum_{i=1}^{m}{P_i^\text{s}}})\right)F_{g^\text{sd}}(\frac{e^R-1}{\sum_{i=1}^{m}{P_i^\text{s}}})\nonumber\\&\,\,\,\,\,\,\,\,\,=\left(e^{-\lambda^\text{sr}\frac{e^R-1}{\sum_{i=1}^{m}{P_i^\text{s}}}}-e^{-\lambda^\text{sr}\frac{e^R-1}{\sum_{i=1}^{m-1}{P_i^\text{s}}}}\right)(1-e^{-\lambda^\text{sd}\frac{e^R-1}{\sum_{i=1}^{m}{P_i^\text{s}}}}),\nonumber
\\
&\beta_m=\left(F_{g^\text{sd}}(\frac{e^R-1}{\sum_{i=1}^{m-1}{P_i^\text{s}}})-F_{g^\text{sd}}(\frac{e^R-1}{\sum_{i=1}^{m}{P_i^\text{s}}})\right)F_{g^\text{sr}}(\frac{e^R-1}{\sum_{i=1}^{m-1}{P_i^\text{s}}})\nonumber\\&\,\,\,\,\,\,\,\,=\left(e^{-\lambda^\text{sd}\frac{e^R-1}{\sum_{i=1}^{m}{P_i^\text{s}}}}-e^{-\lambda^\text{sd}\frac{e^R-1}{\sum_{i=1}^{m-1}{P_i^\text{s}}}}\right)(1-e^{-\lambda^\text{sr}\frac{e^R-1}{\sum_{i=1}^{m-1}{P_i^\text{s}}}}),\nonumber
\\
&\gamma_M=\left(1-e^{-\lambda^\text{sd}\frac{e^R-1}{\sum_{i=1}^{M}{P_i^\text{s}}}}\right)\left(1-e^{-\lambda^\text{sr}\frac{e^R-1}{\sum_{i=1}^{M}{P_i^\text{s}}}}\right).
\end{align}
With the same procedure, the probability that the destination decodes the codeword at the $m$-th round (and not before), i.e., $\Pr(A_m),$ is obtained by
\vspace{-0mm}
\begin{align}\label{eq:RTDprobAm}
&\Pr(A_m)=\beta_m+\sum_{j=1}^{m-1}{\varepsilon_{j,m}},\nonumber
\\
&\varepsilon_{j,m}=\Pr\bigg(  \log(1+g^\text{sr}\sum_{i=1}^{j-1}{P_i^\text{s}})<R\le \log(1+g^\text{sr}\sum_{i=1}^{j}{P_i^\text{s}}) \,\cap   \nonumber\\&\,\,\,\,\,\,\,\,\,\,\,\,\,\,\,\,\,\,\,\,\,\,\,\,\,\,\,\,\,\,\,\, \log(1+g^\text{sd}\sum_{i=1}^{j}{P_i^\text{s}}+g^\text{rd}\sum_{i=j+1}^{m-1}{P_i^\text{r}})<R\nonumber\\&\,\,\,\,\,\,\,\,\,\,\,\,\,\,\,\,\,\,\,\,\,\,\,\,\,\,\,\,\,\,\,\,\,\le \log(1+g^\text{sd}\sum_{i=1}^{j}{P_i^\text{s}}+g^\text{rd}\sum_{i=j+1}^{m}{P_i^\text{r}})                         \bigg).
\end{align}
Here, $\varepsilon_{j,m}$ is the probability that the relay decodes the codeword at the $j$-th round and helps the destination until it decodes the message at round $m,\,j<m$. Thus, (\ref{eq:RTDprobAm}) gives the message decoding probability of destination for all possible activation conditions of the relay. For independent Rayleigh-fading condition, $\varepsilon_{j,m}$ is found as
\vspace{-2mm}
\begin{align}\label{eq:varepsiloncal}
\begin{array}{l}
\varepsilon_{j,m}=\omega_j\theta_{j,m},
\\
\omega_j=F_{g^\text{sr}}(\frac{e^R-1}{\sum_{i=1}^{j-1}{P_i^\text{s}}})-F_{g^\text{sr}}(\frac{e^R-1}{\sum_{i=1}^{j}{P_i^\text{s}}})\\\,\,\,\,\,\,\,\,=e^{-\lambda^\text{sr}\frac{e^R-1}{\sum_{i=1}^{j}{P_i^\text{s}}}}-e^{-\lambda^\text{sr}\frac{e^R-1}{\sum_{i=1}^{j-1}{P_i^\text{s}}}},
\\
\theta_{j,m}=\int_0^{\frac{e^R-1}{\sum_{i=1}^j{P_i^\text{s}}}}{{  f_{g^\text{sd}}(x)}}\times\\{{\Pr\left(g^\text{rd}\sum_{i=j+1}^{m-1}{P_i^\text{r}}< e^R-1-\sum_{i=1}^{j}{P_i^\text{s}}x  \le g^\text{rd}\sum_{i=j+1}^{m}{P_i^\text{r}}   \right) }\text{d}x}
\\=\int_0^{\frac{e^R-1}{\sum_{i=1}^j{P_i^\text{s}}}}{ \lambda^\text{sd}e^{-\lambda^\text{sd}x}\times}\\\,\,\,\,\,\,\,\,\,\,\,\,\,\,\,\,\,\,\,\,{\bigg(e^{-\lambda^\text{rd}(\frac{e^R-1-\sum_{i=1}^{j}{P_i^\text{s}}x}{\sum_{i=j+1}^{m}{P_i^\text{r}}})}-   e^{-\lambda^\text{rd}(\frac{e^R-1-\sum_{i=1}^{j}{P_i^\text{s}}x}{\sum_{i=j+1}^{m-1}{P_i^\text{r}}})}        \bigg)   \text{d}x}\\\,\,\,\,\,\,\,\,\,\,\,=\mathcal{G}_{j}(\sum_{i=j+1}^{m-1}{P_i^\text{r}})- \mathcal{G}_{j}(\sum_{i=j+1}^{m}{P_i^\text{r}}),\\\mathcal{G}_{j}(x)\doteq \\\begin{cases}
{1-e^{-\frac{\lambda^\text{sd}(e^R-1)}{\sum_{i=1}^j{P_i^\text{s}}}}}\\{-\frac{e^{-\frac{\lambda^\text{rd}(e^R-1)}{x}}-e^{-\frac{\lambda^\text{sd}(e^R-1)}{\sum_{i=1}^j{P_i^\text{s}}}}}{1-\frac{\lambda^\text{rd}\sum_{i=1}^j{P_i^\text{s}}}{\lambda^\text{sd}x}}}, & \text{ if } x\ne \frac{\lambda^\text{rd}\sum_{i=1}^j{P_i^\text{s}}}{\lambda^\text{sd}} \\
{1-e^{-\frac{\lambda^\text{sd}(e^R-1)}{\sum_{i=1}^j{P_i^\text{s}}}}}\\{-\frac{\lambda^\text{sd}(e^R-1)}{\sum_{i=1}^j{P_i^\text{s}}}e^{-\frac{\lambda^\text{rd}(e^R-1)}{x}}}, & \text{ if } x= \frac{\lambda^\text{rd}\sum_{i=1}^j{P_i^\text{s}}}{\lambda^\text{sd}}
\end{cases}
 \end{array}
\end{align}
where $\omega_j$ is the decoding probability of the relay at round $j$ (and not before). Also, $\theta_{j,m}$ represents the decoding probability of the destination at the $m$-th round, given that the relay is active in rounds $j+1,\ldots,m.$

Finally, $\Pr(B_{n,m}),$ i.e., the probability that the relay is active in rounds $n+1,\ldots,m,$ is determined as
\begin{align}\label{eq:RTDprobBnm}
&\Pr(B_{n,m}) = \left\{ \begin{array}{l}
 {\varepsilon_{n,m}}\,\,\,\,\,\,\,\,\text{if}\,\,m\ne M+1, \\
 \vartheta_{n} \,\,\,\,\,\,\,\,\,\,\,\,\text{if}\,\,m=M+1, \\
 \end{array} \right.\nonumber
\\
&\vartheta_{n}=\Pr\bigg(\log(1+g^\text{sr}\sum_{i=1}^{n-1}{P_i^\text{s}})<R\le \log(1+g^\text{sr}\sum_{i=1}^{n}{P_i^\text{s}})\nonumber\\&\,\,\,\,\,\,\,\,\,\,\,\,\,\cap\,\log(1+g^\text{sd}\sum_{i=1}^{n}{P_i^\text{s}}+g^\text{rd}\sum_{k=n+1}^{M}{P_k^\text{r}})<R\bigg)=\omega_n\rho_n,
\nonumber\\
&\rho_n=\int_0^{\frac{e^R-1}{\sum_{i=1}^n{P_i^\text{s}}}}{{  f_{g^\text{sd}}(x)F_{g^\text{rd}}\left(\frac{e^R-1-\sum_{i=1}^n{P_i^\text{s}}x}{\sum_{k=n+1}^M{P_k^\text{r}}}\right) }\text{d}x}
\nonumber\\&\,\,\,\,\,\,=\mathcal{G}_n(\sum_{i=n+1}^{M}{P_i^\text{r}}),
\end{align}
where ${\varepsilon_{n,m}}$ and $\mathcal{G}_n(x)$ are defined in (\ref{eq:RTDprobAm}) and (\ref{eq:varepsiloncal}), respectively, and $\rho_n$ is obtained with the same procedure as in (\ref{eq:varepsiloncal}).

Using (\ref{eq:RTDSm})-(\ref{eq:RTDprobBnm}), we can express the outage probability, the throughput and the average power functions of the relay-RTD scheme in terms of $R_{(m)},P_m^\text{s},P_m^\text{r},\,m=1,\ldots,M+1,$ and investigate the system performance, as stated in the following.
\subsection{INR Protocol in Quasi-Static Conditions}
Considering a maximum of $M+1$ INR-based retransmission rounds, $Q$ information nats is encoded into a \emph{parent} codeword of length $l_{(M+1)}=\sum_{m=1}^{M+1}{l_m}.$ Then, the codeword is punctured into $M+1$ subcodewords of lengths $l_{m},\,m=1,\ldots,M+1,$ which are sent by the source/relay in the successive retransmission rounds. In each round, all received subcodewords are combined by the receivers (relay and destination), to decode the message. In this case, the results of \cite[chapter 15]{excellentref,4444444444}, \cite[chapter 7]{ELGAMAL} can be used to show that the maximum data rates which are decodable by the relay and the destination at the $m$-th round are obtained by
\begin{align}\label{eq:functionINRR}
U_m^\text{r}(g^\text{sr})&=\sum_{i=1}^m{\frac{l_i\log(1+g^\text{sr}P_i^\text{s})}{\sum_{k=1}^m{l_k}}}\nonumber\\&=R_{(m)}\sum_{i=1}^m{(\frac{1}{R_{(i)}}-\frac{1}{R_{(i-1)}})\log(1+g^\text{sr}P_i^\text{s})}, \end{align}
and
\begin{align}\label{eq:functionINRD}
&U_{j,m}^\text{d}(g^\text{sd},g^\text{rd})\nonumber\\&=\frac{\sum_{i=1}^j{l_i\log(1+g^\text{sd}P_i^\text{s})}+\sum_{i=j+1}^m{l_i\log(1+g^\text{rd}P_i^\text{r})}}{{\sum_{k=1}^m{l_k}}}\nonumber\\&=R_{(m)}\Big(  \sum_{i=1}^j{(\frac{1}{R_{(i)}}-\frac{1}{R_{(i-1)}})\log(1+g^\text{sd}P_i^\text{s})} \nonumber\\& +  \sum_{i=j+1}^m{(\frac{1}{R_{(i)}}-\frac{1}{R_{(i-1)}})\log(1+g^\text{rd}P_i^\text{r})}        \Big),\, j<m,
\end{align}
respectively, where (\ref{eq:functionINRD}) is based on the assumption that the relay is active in rounds $j+1,\ldots,m.$ Also,
\begin{align}\label{eq:functionINRD2}
U_{m,m}^\text{d}(g^\text{sd})\doteq&\frac{\sum_{i=1}^m{l_i\log(1+g^\text{sd}P_i^\text{s})}}{{\sum_{k=1}^m{l_k}}}\nonumber\\&=R_{(m)}  \sum_{i=1}^m{(\frac{1}{R_{(i)}}-\frac{1}{R_{(i-1)}})\log(1+g^\text{sd}P_i^\text{s})}
\end{align}
denotes the maximum decodable rate of the destination at the $m$-th round, given that the relay is inactive.

Although having high throughput and low outage probability, variable-length coding INR results in high
\emph{packeting} complexity \cite{ARQGlarsson,Tcomkhodemun}. In order to reduce the complexity, fixed-length coding INR scheme is normally considered where setting $l_{m}=l,\forall m,$ in (\ref{eq:functionINRR})-(\ref{eq:functionINRD2}) leads to $R_{(m)}=\frac{R}{m}$ and
\begin{align}\label{eq:Ufixedlength}
U_m^\text{r, fixed-length}(g^\text{sr})=\frac{1}{m}\sum_{i=1}^m{\log(1+g^\text{sr}P_i^\text{s})},
\end{align}
\begin{align}\label{eq:UUfixedlength}
&U_{j,m}^\text{d, fixed-length}(g^\text{sd},g^\text{rd})\nonumber\\&=\frac{1}{m}\left({\sum_{i=1}^j{\log(1+g^\text{sd}P_i^\text{s})}+\sum_{i=j+1}^m{\log(1+g^\text{rd}P_i^\text{r})}}\right),j\le m.
\end{align}

From (\ref{eq:functionINRR})-(\ref{eq:UUfixedlength}), we can find the probabilities $\Pr(A_m),\,\Pr(B_{n,m}),\,\Pr(S_m)$ for the INR protocol; Replacing the terms, e.g., $\log(1+g^\text{sr}\sum_{i=1}^{m}{P_i^\text{s}})$ of the RTD by $R_{(m)}\sum_{i=1}^m{(\frac{1}{R_{(i)}}-\frac{1}{R_{(i-1)}})\log(1+g^\text{sr}P_i^\text{s})}$ for the INR, one can use the same procedure as in (\ref{eq:RTDSm})-(\ref{eq:RTDprobBnm}) to recalculate the parameters $\alpha_m,$ $\beta_m,$ $\gamma_M,$ $\omega_j,$ $\theta_{j,m},$ $\rho_n$ and, consequently, $\Pr(A_m),\,\Pr(B_{n,m}),\,\Pr(S_m)$ for the INR. The details are presented in the Appendix where the probabilities are determined by obtaining $\Pr(U_m^\text{r}(g^\text{sr})\le R_{(m)})$ and $\Pr(U_{j,m}^\text{d}(g^\text{sd},g^\text{rd})\le R_{(m)}),\forall j,m$.

As explained in the Appendix, two-dimensional numerical integrations should be used to determine the probability terms $\rho_n$ and $\theta_{j,m}$ for the INR, which are difficult to find. This is particularly because the boundaries of the two-dimensional integrals can not be expressed in closed-form. For this reason, upper bounds of $\rho_n$ and $\theta_{j,m}$ are presented in the Appendix which are tight at low SNRs.  Moreover, Theorems 1-2 provide other approximation methods which simplify the performance analysis in the presence of the INR protocol. The good point in Theorems 1-2 is that the two-dimensional integrals with unknown integration boundaries are replaced by either closed-form expressions or one-dimensional integrals having known boundaries. As a result, the terms $\theta_{j,m}$ and $\rho_n$ can be determined easily.

\emph{\textbf{Theorem 1}:} For the fixed-length INR protocol, the performance of the relay-ARQ setup is underestimated, i.e., the throughput is lower bounded and the  outage probability is upper bounded, via the following inequalities
\begin{align}\label{eq:theoremstate1}
\Pr\left(U_m^\text{r, fixed-length}(g^\text{sr})\le \frac{R}{m}\right)&\le F_{g^\text{sr}}(\frac{e^{\frac{R}{m}}-1}{\sqrt[m]{\prod_{i=1}^m{P_i^\text{s}}}})\nonumber\\&=1-e^{-\lambda^\text{sr}\frac{e^{\frac{R}{m}}-1}{\sqrt[m]{\prod_{i=1}^m{P_i^\text{s}}}}},
\end{align}
\begin{align}\label{eq:theoremstate2}
\Pr&\left(U_{j,m}^\text{d, fixed-length}(g^\text{sd},g^\text{rd})<\frac{R}{m}\right)\nonumber\\&\le 1-V_{j,m}\left((\frac{e^{\frac{R}{m}}-1}{\sqrt[m]{\prod_{i=1}^j{P_i^\text{s}}\prod_{i=j+1}^m{P_i^\text{r}}}})^\frac{m}{m-j}\right),
\end{align}
where $V_{j,m}(v)\doteq \lambda^\text{sd}\int_0^\infty{e^{-\lambda^\text{sd}x-(\lambda^\text{rd}x^\frac{j}{j-m})v}\text{d}x}.$
\begin{proof}
Please see the Appendix.
\end{proof}
Due to properties of the Minkowski's inequality, the bounds are tight at high SNRs. Finally,
we close the discussions with the following theorem which provides an upper-estimate of the system performance.

\emph{\textbf{Theorem 2}:} For sufficiently low SNRs, the performance of the relay-INR protocol is upper-estimated via the following inequalities
\begin{align}\label{eq:theoremstate21}
\Pr\left(U_m^\text{r, fixed-length}(g^\text{sr})\le \frac{R}{m}\right)&\ge F_{g^\text{sr}}\Big(\sqrt{\frac{e^\frac{2R}{m}}{\sqrt[m]{\prod_{i=1}^m{(1+(P_i^\text{s})^2)}}}-1}\Big)\nonumber\\&=1-e^{-\lambda^\text{sr}\sqrt{\frac{e^\frac{2R}{m}}{\sqrt[m]{\prod_{i=1}^m{(1+(P_i^\text{s})^2)}}}-1}},
\end{align}
\begin{align}\label{eq:theoremstate22}
&\Pr\left(U_{j,m}^\text{d, fixed-length}(g^\text{sd},g^\text{rd})<\frac{R}{m}\right)\ge W_{j,m}(r), \nonumber\\&W_{j,m}(r)\doteq\int_0^{\sqrt{\sqrt[j]{r}-1}}{\lambda^\text{sd}e^{-\lambda^\text{sd}x}{\Big(1-e^{-\lambda^\text{rd}\sqrt{\sqrt[m-j]{r}(1+x^2)^\frac{j}{j-m}-1}}\Big)}\text{d}x},\,
\nonumber\\&r=\frac{e^{2R}}{\prod_{i=1}^j{(1+(P_i^\text{s})^2)}\prod_{i=j+1}^m{(1+(P_i^\text{r})^2)}}.
\end{align}
\begin{proof}
Please see the Appendix.
\end{proof}

The tightness of the bounds is verified in Subsection IV.D.
\subsection{Performance Analysis in Fast-Fading Conditions}
Considering the fast-fading models\footnote{Under fast-fading channel conditions, the INR protocol is studied with fixed-length coding because the length of the codewords is the same as the fading block length.}, e.g., \cite{6512535,1661837,ARQGlarsson}, the outage probability, the throughput and the average power functions, i.e., (\ref{eq:throughputdef})-(\ref{eq:outageprob}), are obtained with the same procedure as before while the achievable rate terms, e.g., (\ref{eq:UUfixedlength}) for the INR, are replaced by
\begin{align}\label{eq:UUfastfading}
&U_{j,m}^\text{d, fast-fading}\nonumber\\&=\frac{1}{m}\left({\sum_{i=1}^j{\log(1+g_i^\text{sd}P_i^\text{s})}+\sum_{i=j+1}^m{\log(1+g_i^\text{rd}P_i^\text{r})}}\right),j\le m.
\end{align}
Here, $g_i^\text{sd}$ and $g_i^\text{rd}$ are the source-destination and the relay-destination channel realizations at the $i$-th round. Changing the achievable rate terms, and recalculating the probabilities, is the only modification required for the fast-fading condition and the rest of the procedure does not need to be changed. Specifically, Theorem 3 provides a method for calculating the probabilities in the fast-fading conditions.

\textbf{\emph{Theorem 3}}: The power-limited throughput/outage probability of a relay-ARQ setup utilizing RTD and INR protocols can be analyzed via the equalities
\begin{align}\label{eq:theoremeq31}
&\Pr(\log(1+\sum_{i=1}^j{P_i^\text{s}g_i^\text{sd}}+\sum_{i=j+1}^m{P_i^\text{r}g_i^\text{rd}})\le R)\nonumber\\&=\mathcal{O}_{j,m}(e^R-1)-\mathcal{O}_{j,m}(0),
\nonumber\\&\mathcal{O}_{j,m}(x)\doteq\sum_{i=1}^j{{\frac{e^{-\frac{\lambda^\text{sd}x}{P_i^\text{s}}}}{\prod_{k=1,k\ne i}^j{(1-\frac{P_k^\text{s}}{P_i^\text{s}})}\prod_{k=j+1}^m{(1-\frac{P_k^\text{r}\lambda^\text{sd}}{\lambda^\text{rd}P_i^\text{s}})}}}}\nonumber\\&+\sum_{i=j+1}^m{{\frac{e^{-\frac{\lambda^\text{rd}x}{P_i^\text{r}}}}{\prod_{k=1}^j{(1-\frac{P_k^\text{s}\lambda^\text{rd}}{\lambda^\text{sd}P_i^\text{r}})}\prod_{k=j+1,k\ne i}^m{(1-\frac{P_k^\text{r}}{P_i^\text{r}})}}}},\nonumber\\&P_i^\text{s}\ne P_k^\text{s}, P_i^\text{r}\ne P_k^\text{r},i\ne k,\frac{P_i^\text{s}}{\lambda^\text{sd}}\ne \frac{P_k^\text{r}}{\lambda^\text{rd}},\forall i,k,
\end{align}
and
\vspace{-0mm}
\begin{align}\label{eq:theoremeq32}
&\Pr(\sum_{i=1}^j{\log(1+P_i^\text{s}g_i^\text{sd})}+\sum_{i=j+1}^m{\log(1+P_i^\text{r}g_i^\text{rd})}\le R)=\nonumber\\&1-K_{j,m} \mathcal{H}_{1,m+1}^{m+1,0}\bigg[\frac{ e^{R}}{C_{j,m}}\bigg |_{(0,1,0),(1,1,\frac{\lambda^\text{sd}}{P_1^\text{s}},1),\ldots,(1,1,\frac{\lambda^\text{sd}}{P_j^\text{s}},1),}^{\,\,\,\,\,\,\,\,\,\,\,\,\,\,\,\,\,\,\,\,\,\,\,\,\,\,\,\,\,\,\,\,\,\,\,\,\,\,\,\,\,\,\,\,\,\,(1,1,0)} \nonumber\\&\,\,\,\,\,\,\,\,\,\,\,\,\,\,\,\,\,\,\,\,\,\,\,\,\,\,\,\,\,\,\,\,\,\,\,\,\,\,\,\,\,\,\,\,\,\,\,\,\,\,\,\,\,\,\,\,\,\,\,\,\,\,\,\,\,\,\,\,\,\,\,\,\,\,\,\,\,\,\,\,\,\,\,_{(1,1,\frac{\lambda^\text{rd}}{P_{j+1}^\text{r}},1),\ldots,(1,1,\frac{\lambda^\text{rd}}{P_{m}^\text{r}},1)}\bigg]\nonumber\\&K_{j,m}\doteq e^{(\lambda^\text{sd}\sum_{i=1}^j{\frac{1}{P_i^\text{s}}}+\lambda^\text{rd}\sum_{i=j+1}^m{\frac{1}{P_i^\text{r}}})},\, \nonumber\\&C_{j,m}\doteq\frac{1}{(\lambda^\text{sd})^j (\lambda^\text{rd})^{m-j}}(\prod_{i=1}^j{P_i^\text{s}})(\prod_{i=j+1}^m{P_i^\text{r}}),
\end{align}
respectively, if the channel is fast-fading.
Here, $ \mathcal{H}_{m,n}^{v,w}\big[. \big]$ denotes the generalized upper incomplete Fox'H function \cite{5357980}.
\begin{proof}
Please see the Appendix.
\end{proof}
Finally, to enjoy the practical benefits of the relay-ARQ the channel code should satisfy the following requirements:
\begin{itemize}
  \item[1)] A parent code that can be punctured into rate-optimized subcodewords and
  \item[2)] decoders at the relay/destination with performance close to (\ref{eq:RTDSm})-(\ref{eq:functionINRD2}), (\ref{eq:UUfastfading}) for all retransmissions.
\end{itemize}
There exist several practical code designs, e.g., \cite{revisiontvtrelay1,1327848}, that satisfy these requirements. Moreover, to implement adaptive power allocation, the source and the relay should be equipped with adaptive power amplifiers. However, as the power adaptation is based on the long-term channel statistics with finite levels of transmission powers, the power amplifiers can be efficiently designed.
\vspace{-0mm}
\subsection{Simulations and Discussions}
Using the same arguments as in \cite{Tcomkhodemun,6566132,5771499}, it can be showed that both the power-limited throughput maximization and the outage probability minimization of ARQ protocols are nonconvex optimization problems, even for the single-user (without relay) setup. Therefore, the problems should be solved via iterative optimization algorithms. In our simulations, the number of optimization parameters is low enough to use exhaustive search, which is what we have used for our simulations. Also, for faster convergence, we have repeated the simulations by using the iterative algorithm of \cite{Tcomkhodemun}, and by using ``fminsearch'' and ``fmincon'' functions of MATLAB. Using the closed-form expressions, the results have been obtained for different initial settings and we have tested the fmincon function for ``\emph{interior-point},'' ``\emph{active-set}'' and ``\emph{trust-region-reflective}'' options of the optimization algorithm. In all cases, the results are the same with high accuracy, which is an indication of a reliable result.
Finally, note that, except for Fig. 6 where we verify the tightness of the bounds in Theorems 1-2, the results of the figures are obtained both analytically and via Monte Carlo simulations which lead to the same results. Thus, Figs. 2-5, 7-15 represent both the analytical and the Monte Carlo-based simulation results.

In the simulations, fixed-length coding is considered for the INR protocol, unless otherwise stated. Also, in all figures, except Figs. 12-13, the results are obtained for the quasi-static conditions. The fast-fading case is considered in Figs. 12-13.

\emph{Performance analysis with individual power constraints on the source and the relay:} In Figs. 2-3, we study the system performance in scenario 2, where there are individual power constraints on the source and the relay. The results show that 1) with individual power constraints, optimal power allocation increases the throughput at low SNRs. However,  the gain of optimal power allocation is negligible, when the goal is to maximize the throughput at high SNRs (Fig. 2).  2) Considerable outage probability reduction is achieved by optimal power allocation, particularly at high SNRs (Fig. 3). Finally, 3) increasing the maximum number of retransmissions leads to marginal throughput increment, especially at high SNRs, when the source and the relay are individually power-limited (Fig. 2).

\begin{figure}\label{fig:NPA}
\vspace{-5mm}
\centering
  \includegraphics[width=.98\columnwidth]{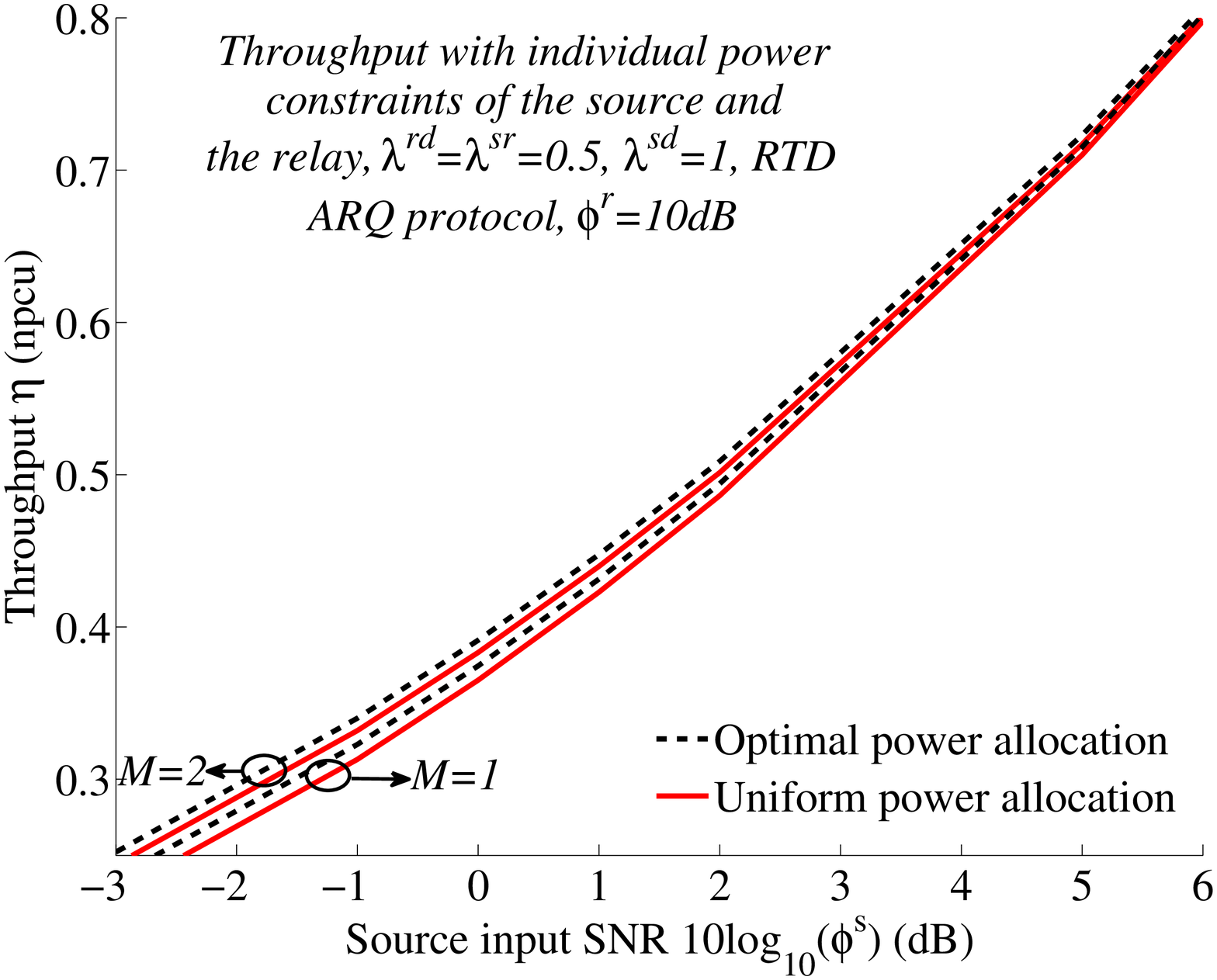}\\\vspace{-0mm}
\caption{Throughput vs the average transmission SNR of the source $10\log_{10}(\phi^\text{s})$. The results are obtained with individual power constraints on the source and the relay (scenario 2) with $\Phi^\text{r}=10\text{dB}.$ RTD ARQ protocol, $\lambda^\text{sr}=\lambda^\text{rd}=0.5, \lambda^\text{sd}=1.$}\label{figure111}
\vspace{-0mm}
\end{figure}
\begin{figure}\label{fig:NPA}
\vspace{-0mm}
\centering
  \includegraphics[width=.98\columnwidth]{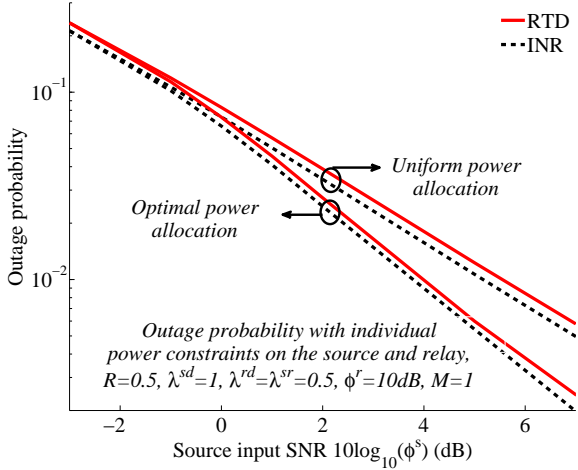}\\\vspace{-0mm}
\caption{Outage probability vs the average transmission SNR of the source $10\log_{10}(\phi^\text{s})$. The results are obtained with individual power constraints on the source and the relay (scenario 2) with $\Phi^\text{r}=10\text{dB}.$ Initial transmission rate $R=0.5$, $M=1$, $\lambda^\text{sr}=\lambda^\text{rd}=0.5, \lambda^\text{sd}=1.$}\label{figure111}
\vspace{-0mm}
\end{figure}

Figures 4-15 present the simulation results for the case with a sum power constraint on the source and the relay, i.e., $\Phi^\text{total}\le\phi^\text{total}$ in (\ref{eq:optproblem}), where $10\log_{10}(\phi^\text{total})$ denotes the input SNR.

\emph{On the coverage region of the relay-ARQ network:} Let us define the \emph{coverage region} as
\begin{align}\label{eq:coveragedefARQ}
\mathcal{R}(\epsilon|\lambda^\text{rd},\lambda^\text{sr},M)\doteq \{\lambda^\text{sd}|\lambda^\text{rd},\lambda^\text{sr}, M, \Pr(\text{Outage})\le \epsilon\}.
\end{align}
That is, (\ref{eq:coveragedefARQ}) defines the set of fading coefficients $\lambda^\text{sd}$  (for a given SNR, $M,$ $\lambda^\text{rd},$ and $\lambda^\text{sr}$) which leads to $\Pr(\text{Outage})\le \epsilon.$ Fig. 4 demonstrates the \emph{relative} coverage region of different communication setups, compared to the single-user setup without ARQ, i.e., $\varphi=\frac{\mathcal{R}(\epsilon|\lambda^\text{rd},\lambda^\text{sr},M)}{\mathcal{R}(\epsilon|\infty,\infty,0)}$. In other words, each curve in Fig. 4 shows the gain of the considered scheme compared to the single-user system without ARQ. 
The higher the curve is, the wider the coverage region is\footnote{Note that the variances of the fading coefficients, that can quantify the distances between the transmission endpoints, are given by, e.g., $\frac{1}{\lambda^\text{sd}}.$}.
The results show that, compared to the single-user setup, the implementation of the relays leads to considerable coverage region increment. Also, compared to uniform power allocation (which corresponds to, e.g., \cite{milicomARQrelay,elsevierARQ2007,a06247450,a06188994,a06362532,a06189805}), our proposed power-optimized scheme increases the coverage region of the relay-assisted networks substantially. For instance, consider the INR ARQ, $M=1$ and the coverage threshold $\epsilon=10^{-3}.$ In this case, the implementation of power-optimized ARQ in single-user and relay-assisted setups increases the coverage region by 17 and 27 times, respectively. With uniform power allocation, there is a (almost) fixed gap between the performance of the INR and RTD protocols. With optimal power allocation, the difference between the coverage regions of the RTD- and INR-based schemes decreases.

\begin{figure}\label{fig:NPA}
\vspace{-0mm}
\centering
  \includegraphics[width=.98\columnwidth]{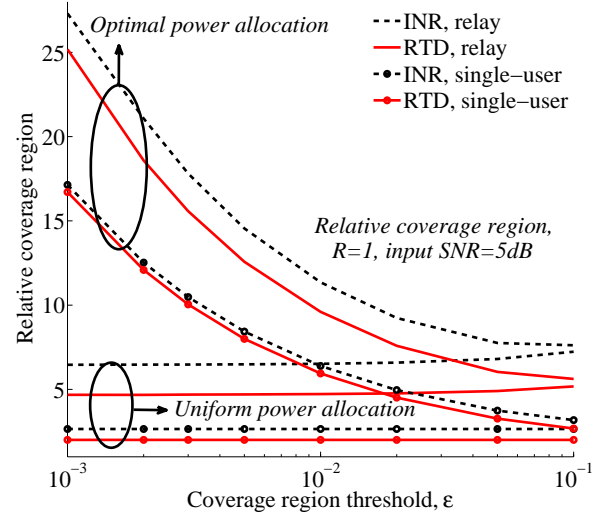}\\\vspace{-0mm}
\caption{Relative coverage region, compared to the single-user (without relay) setup without ARQ. Scenario 1 (sum power constraint $\Phi^\text{total}\le\phi^\text{total}$ in (\ref{eq:optproblem}), with $10\log_{10}(\phi^\text{total})$ being the input SNR). $M=1,$ initial transmission rate $R=1.$ For relay-assisted channel, $\lambda^\text{sr}=\lambda^\text{rd}=0.5.$}\label{figure111}
\vspace{-0mm}
\end{figure}
\begin{figure}\label{fig:NPA}
\vspace{-0mm}
\centering
  \includegraphics[width=.98\columnwidth]{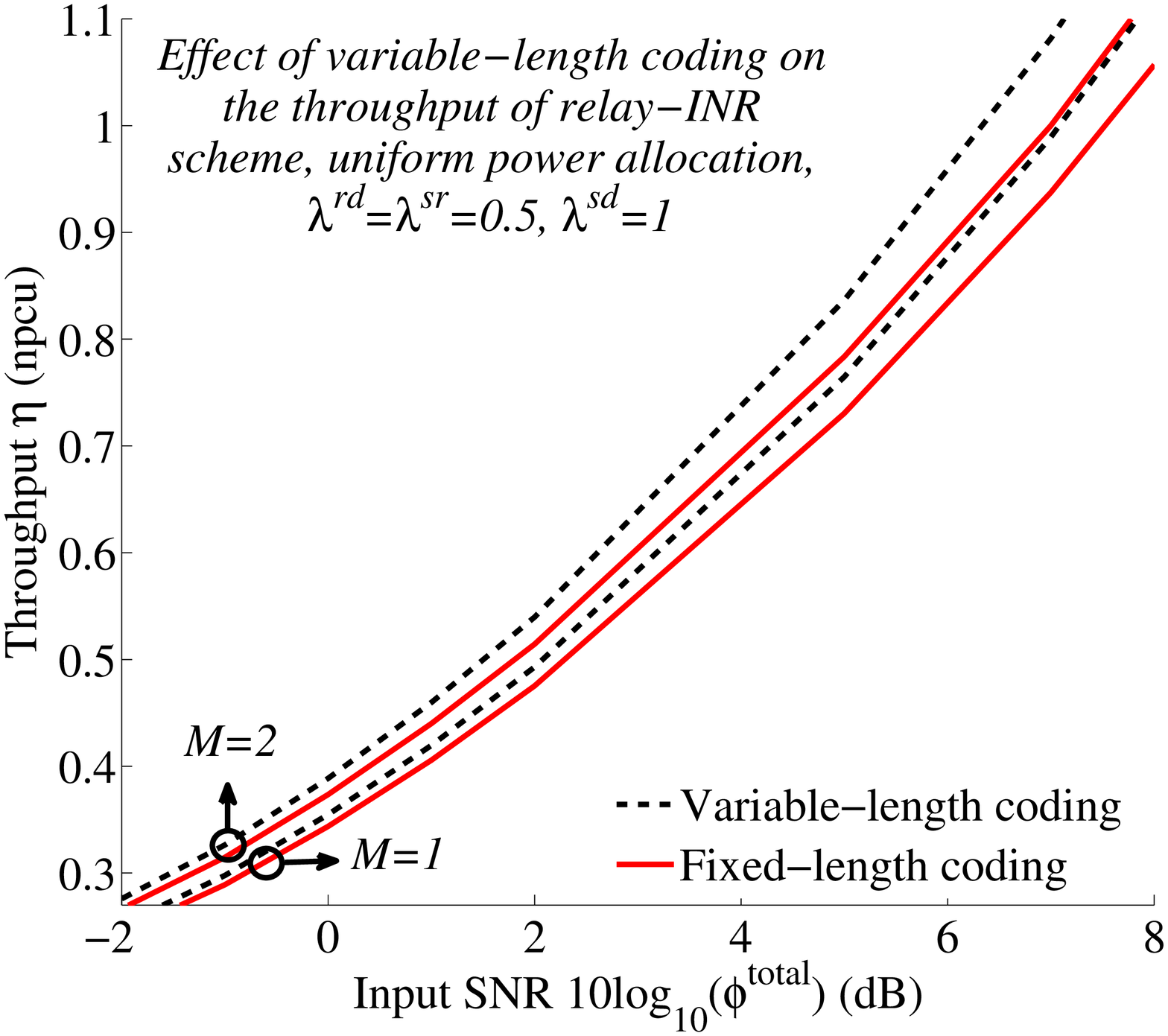}\\\vspace{-0mm}
\caption{Effect of variable-length coding on the throughput of the relay-INR protocol, uniform power allocation, $\lambda^\text{sr}=\lambda^\text{rd}=0.5,\lambda^\text{sd}=1.$}\label{figure111}
\vspace{-0mm}
\end{figure}
\vspace{-0mm}
\begin{figure}\label{fig:NPA}
\vspace{-0mm}
\centering
  \includegraphics[width=.98\columnwidth]{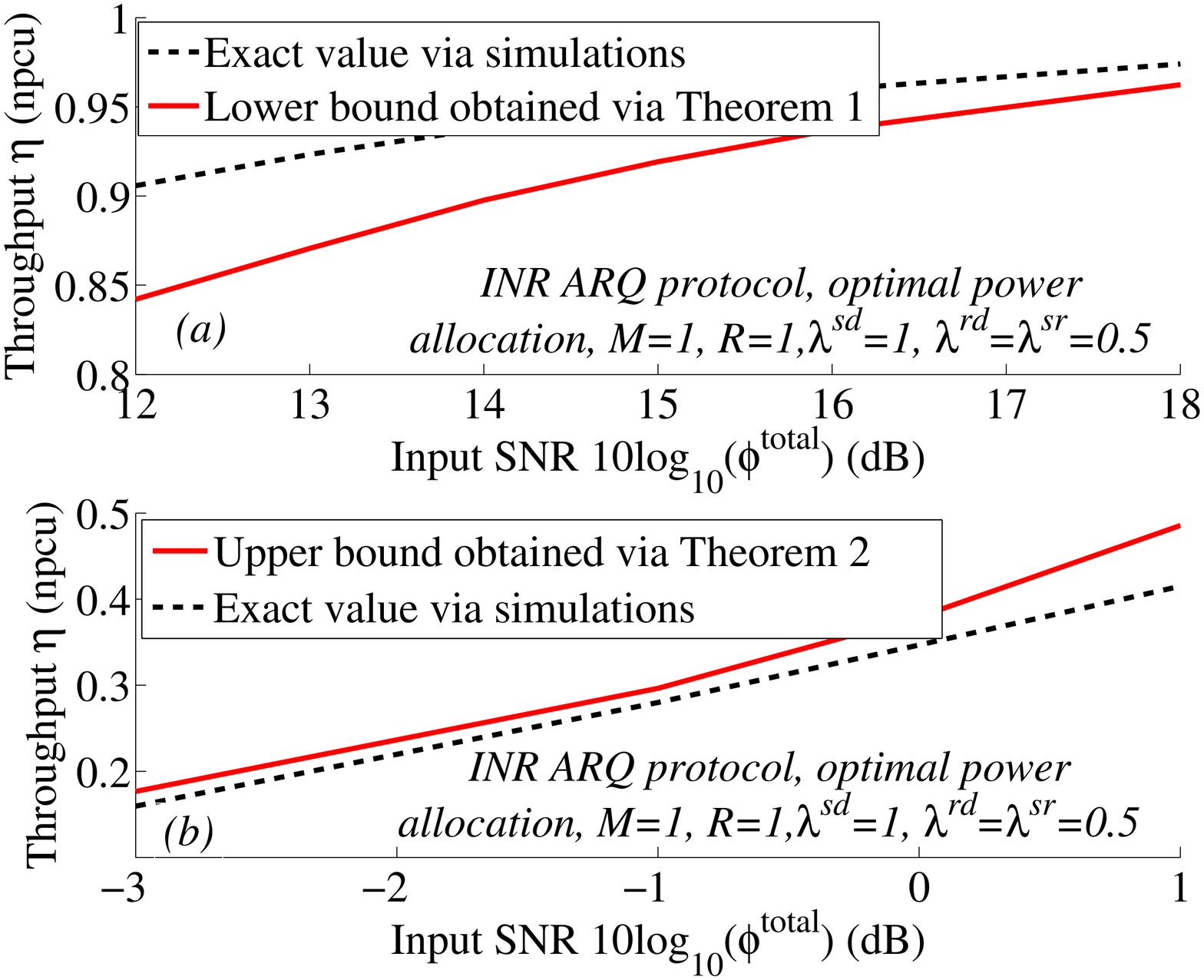}\\\vspace{-0mm}
\caption{On the effect of the bounds introduced in Theorems 1-2, optimal power allocation, scenario 1, $M=1,$ initial transmission rate $R=1,$  $\lambda^\text{sd}=1,$ $\lambda^\text{sr}=\lambda^\text{rd}=0.5.$}\label{figure111}
\vspace{-0mm}
\end{figure}

\emph{On the effect of variable-length coding:} The effect of variable-length coding on the performance of the relay-INR approach is investigated in Fig. 5. Compared to fixed-length coding, considerable (resp. marginal) throughput increment is achieved by variable-length coding at moderate (resp. low) SNRs. Also, the effect of variable-length coding increases with the maximum number of retransmissions, i.e., $M$.

\emph{On the tightness of the proposed bounds:} In Fig. 6, we compare the throughput bounds achieved via Theorems 1-2 with the exact throughput obtained through (\ref{eq:Ufixedlength})-(\ref{eq:UUfixedlength}). The results are obtained with optimal power allocation and for a fixed initial transmission rate $R=1.$ As shown, the bound presented in Theorem 1 (resp. Theorem 2) is tight at high (resp. low) SNR and their tightness decreases as the input SNR decreases (resp. increases).

\emph{Throughput and outage probability in power-optimized relay-ARQ scheme:} Considering the RTD, Fig. 7 demonstrates the throughput versus the outage probability. Moreover, Fig. 8 studies the effect of optimal power allocation on the outage probability.
Considering the figures, it is deduced that 1) with a sum power constraint on the source and the relay, optimal power allocation improves both the throughput and the outage probability considerably. 2) For a given outage probability, increasing the number of retransmissions leads to substantial throughput increment, if the (re)transmission powers are allocated optimally (Fig. 7). Intuitively, this is because with more number of retransmissions the relay gets more involved and the good characteristics of the relay-destination link are properly exploited.
3) The difference between the outage-limited throughput of optimal and uniform power allocation schemes increases with the number of retransmissions and decreases with the outage probability (Fig. 7). 4) The difference between the outage probability of the single-user and relay networks, i.e., the gain of implementing the relay node, increases with the input SNR (Fig. 8).

\emph{Comparison with the state-of-the-art schemes:} In Fig. 9, we compare the data transmission efficiency of the proposed relay-ARQ approach with ones in different data transmission approaches. As demonstrated in the figure, the proposed relay-ARQ approach outperforms the other schemes including 1) the single-user channel without ARQ \cite{Tcomkhodemun}, 2) single-user setup using ARQ \cite{6566132}, 3) the relay channel without ARQ, e.g., \cite{a04133862,a4450823} with one relay, 4) the relay-ARQ in the cases where the source-destination link is ignored, e.g., \cite{a06399140,a06214030} with one relay, or 5) the relay-ARQ when, using STC, the data is simultaneously retransmitted by both the source and the relay in the retransmissions \cite{a06213570,6477555,6512535}. Moreover, the gain of the proposed scheme, compared to the considered state-of-the-art protocols, increases with the SNR.

Here, it is interesting to note that, compared to the single-user setups, the implementation complexity increases in relay-assisted systems; this is because the data transmission is based on more handshakings between the terminals when the relay is utilized. Also, compared to the single-user systems, more feedback resources are required in the relay-based system. On the other hand, the proposed scheme is less complex compared to the simultaneous transmission-based relay-ARQ scheme. This is because the simultaneous transmission schemes, e.g., \cite{a04786513,a06133849,a06051530,a06184258,a05956564,a05992825,a06213570,ARQcollaborative2006,lanati2,ARQcooperativeglobcom,ARQrelaynarasimhan,6477555,6512535}, are based on the assumptions that either different frequency bands are used by the source and the relay or the STC-based techniques are used while the source and the relay are perfectly synchronized. With our data transmission model, none of these assumptions are required.
\begin{figure}\label{fig:NPA}
\vspace{-0mm}
\centering
  \includegraphics[width=.98\columnwidth]{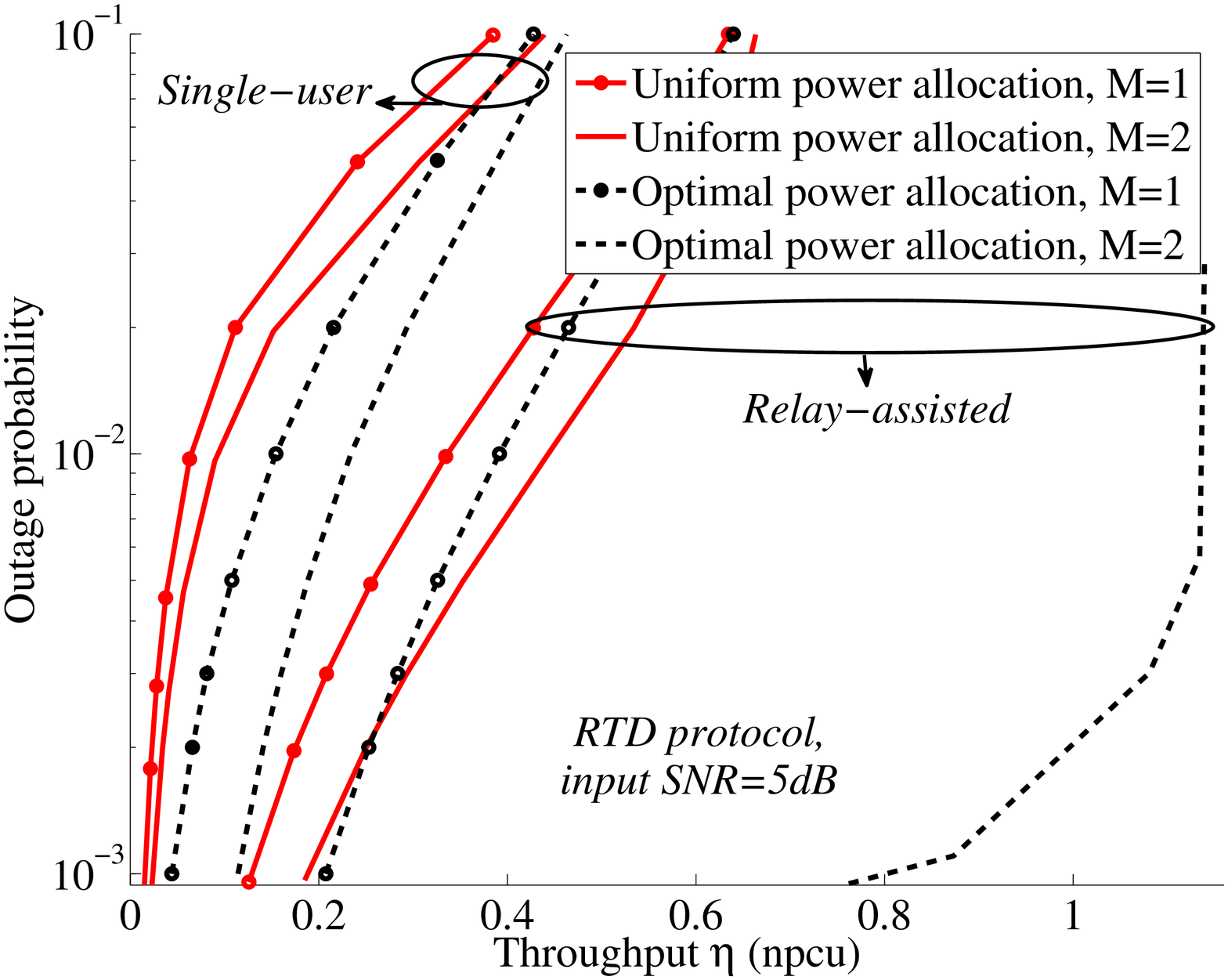}\\\vspace{-0mm}
\caption{Throughput vs the outage probability, RTD ARQ protocol, input SNR $10\log_{10}(\phi^\text{total})=5\text{dB},$ scenario 1, $\lambda^\text{sd}=1,$ $\lambda^\text{sr}=\lambda^\text{rd}=0.5.$ For a given outage probability, the initial transmission rate (and the retransmission powers, if required) are optimized to maximize the throughput.}\label{figure111}
\vspace{-0mm}
\end{figure}
\begin{figure}\label{fig:NPA}
\vspace{-0mm}
\centering
  \includegraphics[width=.98\columnwidth]{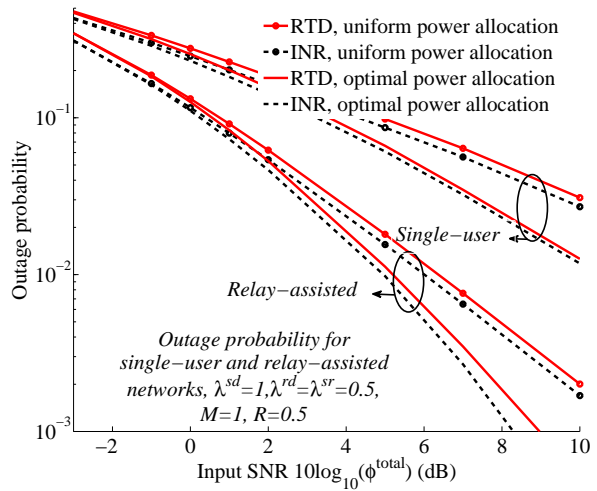}\\\vspace{-0mm}
\caption{Outage probability for single-user and relay-assisted networks, sum power constraint (scenario 1), $M=1,$ initial transmission rate $R=0.5.$ For both single-user and relay-assisted channels, we set $\lambda^\text{sd}=1.$ For relay-assisted channel, $\lambda^\text{sr}=\lambda^\text{rd}=0.5.$}\label{figure111}
\vspace{-0mm}
\end{figure}

\begin{figure}\label{fig:NPA}
\vspace{-0mm}
\centering
  \includegraphics[width=.98\columnwidth]{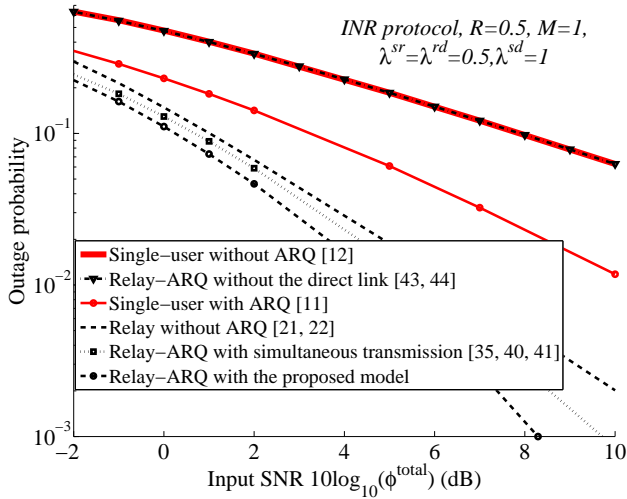}\\\vspace{-0mm}
\caption{Outage probability for different data transmission schemes, INR ARQ protocol, $R=0.5,\,M=1,$ $\lambda^\text{sd}=1,$ $\lambda^\text{sr}=\lambda^\text{rd}=0.5.$}\label{figure111}
\vspace{-0mm}
\end{figure}

\emph{On the optimal power terms:} Fig. 10 shows an example of optimal power terms minimizing the outage probability. Different behaviors are observed at low and high SNRs, which can be interpreted as follows.

At low SNRs, the power resources are limited and the ARQ scheme is \emph{conservative}.
In this case, $P_1^\text{s}$ is set to be high, such that either the relay or the destination can decode the data at the end of round 1 and, as a result, $P_2^\text{s}$ is low. Then, if the relay decodes the message, high power is given to the relay to exploit the good properties of the relay-destination link and increase the probability of successful decoding.

At high SNRs, i.e., when the sum power constraint is relaxed, there is enough power to \emph{gamble}. Here, $P_1^\text{s}$ is set to be low (but high enough such that the relay can decode the message with high probability). If the source-destination channel is \emph{bad}, the data can not be decoded by the destination and is retransmitted with higher powers. On the other hand, if the source-destination channel experiences good conditions, this gambling brings high return. Moreover, we have $ P_2^\text{r}<P_2^\text{s}$ because 1) the relay-destination link experiences better average characteristics than the source-destination link and the relay requires less power than the source to guarantee a given outage probability. Also, 2) with high SNR, it is more likely that the data is retransmitted through the relay than the source. Thus, to keep the average sum power limited, less power is given to the relay than the source in the second round.

\begin{figure}\label{fig:NPA}
\vspace{-0mm}
\centering
  \includegraphics[width=.98\columnwidth]{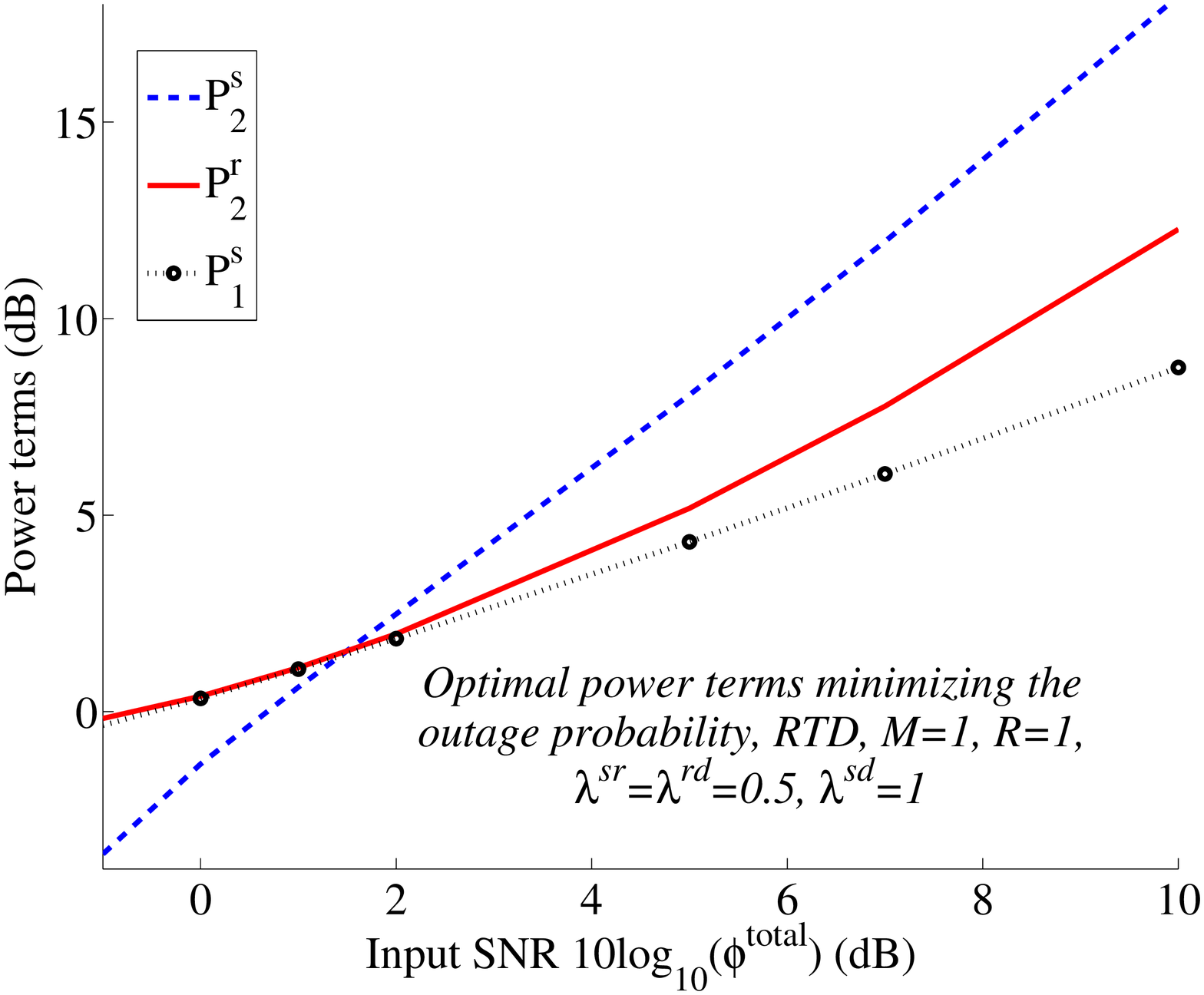}\\\vspace{-0mm}
\caption{Optimal power terms minimizing the outage probability of the relay-RTD network, $R=1,\,M=1,$ $\lambda^\text{sd}=1,$ $\lambda^\text{sr}=\lambda^\text{rd}=0.5.$}\label{figure111}
\vspace{-0mm}
\end{figure}

\emph{On the effect of imperfect feedback signals:} In wireless networks, the feedback signals reach the transmitter through a communication link experiencing different levels of noise and fading. Hence, it is probable to receive erroneous feedback signals at the transmitter(s).
Setting $M=1$, Fig. 11 studies the effect of imperfect feedback channels on the performance of the single-user and relay communication setups using RTD. Here, we have used the fact that, following the same arguments as before and with $M=1$, the total average power and the outage probability of the relay-RTD network are respectively found as
\begin{align}\label{eq:noisypower}
&\Phi^\text{total}=\frac{E\{\Xi^\text{total}\}}{E\{\mathcal{T}^\text{total}\}},\nonumber\\
&E\{\Xi^\text{total}\}=P_1^\text{s}+\big(\omega_1(1-p^\text{sr})P_2^\text{r}+\omega_1p^\text{sr}(P_2^\text{r}+P_2^\text{s})\nonumber\\&+(1-\omega_1)(1-p^\text{sr})P_2^\text{s}\big)\big((1-\beta_1)(1-p^\text{rd})(1-p^\text{sd})+\beta_1p^\text{rd}p^\text{sd}\big) \nonumber\\&+ p^\text{sd}(1-p^\text{rd})\big(P_2^\text{r}\omega_1(1-\beta_1)+P_2^\text{s}\beta_1\big) \nonumber\\&+ p^\text{rd}(1-p^\text{sd})\big(P_2^\text{s}(1-\beta_1)+P_2^\text{r}\omega_1\beta_1\big),\nonumber\\
&E\{\mathcal{T}^\text{total}\}=\nonumber\\&1+\big(1-(1-\omega_1)p^\text{sr}\big)\big((1-\beta_1)(1-p^\text{rd})(1-p^\text{sd})+\beta_1p^\text{rd}p^\text{sd}\big)\nonumber\\&+p^\text{sd}(1-p^\text{rd})\big(\beta_1+(1-\beta_1)\omega_1\big)+p^\text{rd}(1-p^\text{sd})\big(1-\beta_1+\omega_1\beta_1\big),
\end{align}
\begin{align}\label{eq:noisyoutage}
&\Pr(\text{Outage})=(1-\beta_1)p^\text{rd}p^\text{sd}+p^\text{sd}(1-p^\text{rd})\omega_1\mu\nonumber\\&+p^\text{sd}(1-p^\text{rd})(1-\omega_1)(1-\beta_1)+(1-p^\text{sd})p^\text{rd}\sigma\nonumber\\&+(1-p^\text{rd})(1-p^\text{sd})(\omega_1(1-p^\text{sr})\mu+\omega_1p^\text{sr}\kappa\nonumber\\&+(1-\omega_1)(1-p^\text{sr})\sigma+(1-\beta_1)(1-\omega_1)p^\text{sr}),
\nonumber\\&
\mu=\Pr(\log(1+g^\text{sd}P_1^\text{s}+g^\text{rd}P_2^\text{r})<R),
\nonumber\\&
\sigma=\Pr(\log(1+g^\text{sd}(P_1^\text{s}+P_2^\text{s}))<R),
\nonumber\\&
\kappa=\Pr(\log(1+g^\text{sd}P_1^\text{s}+Z)<R),Z=|H^\text{sd}\sqrt{P_2^\text{s}}+H^\text{rd}\sqrt{P_2^\text{r}}|^2,
\end{align}
if the feedback links are noisy. Here, $p^\text{sd},\,p^\text{rd}$ and $p^\text{sr}$ denote the feedback bit error probability in the destination-source, the destination-relay and the relay-source feedback links, respectively. Moreover, the random variable $Z=|H^\text{sd}\sqrt{P_2^\text{s}}+H^\text{rd}\sqrt{P_2^\text{r}}|^2$ in (\ref{eq:noisyoutage}) comes from the fact that, with an erroneous feedback, the relay and the source may retransmit the data simultaneously, for instance, if the ACK bits sent by the destination and the relay are not correctly received by the source and relay at the end of the first round. Note that setting $p^\text{sd}= p^\text{rd}= p^\text{sr}=0$ in (\ref{eq:noisypower})-(\ref{eq:noisyoutage}), the results are changed to the ones in (\ref{eq:outageprob}), (\ref{eq:totalpower}) with $M=1.$

As shown in Fig. 11, the effect of imperfect feedback signals on the outage probability of the single-user and relay-assisted ARQ schemes is negligible, if the feedback bit error probabilities are in the practical range of interest (The same point is valid for the throughput, although not demonstrated in the figure.). However, the erroneous feedback signal affects the system performance at high feedback error probabilities.

\emph{On temporal variations of the fading coefficients:} In Figs. 12-13, we demonstrate the outage probability and the throughput of the relay-ARQ setup in fast-fading conditions and the results are compared with the ones achieved in a quasi-static channel. Compared to the case with a quasi-static channel, better data transmission efficiency is observed in the fast-fading model. This is because with fast-fading more time diversity is exploited by the ARQ protocols. Also, compared to the quasi-static model, the effect of optimal power allocation on improving the outage probability and the throughput of the relay-INR (resp. relay-RTD) setup increases (resp. decreases) when the channel is fast-fading. Moreover, the simulations show that, with fast-fading, the difference between the performance of the single-user and relay-assisted networks decreases slightly. However, there is still remarkable difference between the performances of these two communication setups and implementation of the relay leads to significant performance improvement. Finally, it is worth noting that we have tested the results of the fast-fading condition in many different cases, but since the results follow the same trend as the ones in the quasi-static model, we have chosen not to include them in the paper.

\begin{figure}\label{fig:NPA}
\vspace{-0mm}
\centering
  \includegraphics[width=.98\columnwidth]{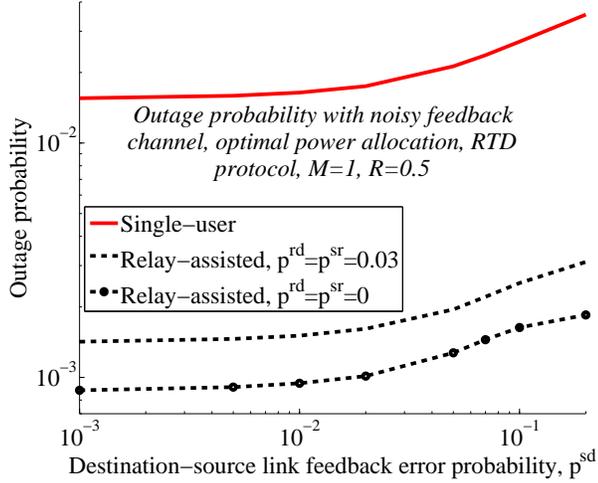}\\\vspace{-0mm}
\caption{On the effect of noisy feedback channel. Optimal power allocation, RTD HARQ protocol, $R=0.5,\,M=1,$ $\lambda^\text{sd}=1, \text{SNR}=10\text{dB}.$ For the relay-assisted setup, we have $\lambda^\text{sr}=\lambda^\text{rd}=0.5.$ }\label{figure111}
\vspace{-0mm}
\end{figure}
\begin{figure}\label{fig:NPA}
\vspace{-0mm}
\centering
  \includegraphics[width=.98\columnwidth]{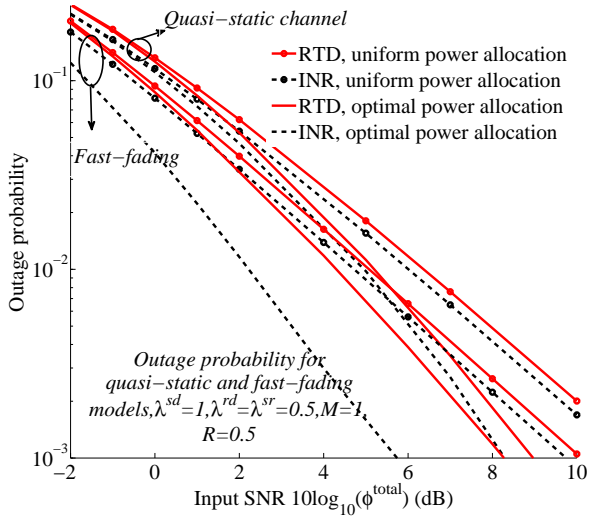}\\\vspace{-0mm}
\caption{The outage probability of the relay-ARQ setup in different fading conditions, $R=0.5,\,M=1,$ $\lambda^\text{sd}=1,$ $\lambda^\text{sr}=\lambda^\text{rd}=0.5.$}\label{figure111}
\vspace{-0mm}
\end{figure}

\begin{figure}\label{fig:NPA}
\vspace{-0mm}
\centering
  \includegraphics[width=.98\columnwidth]{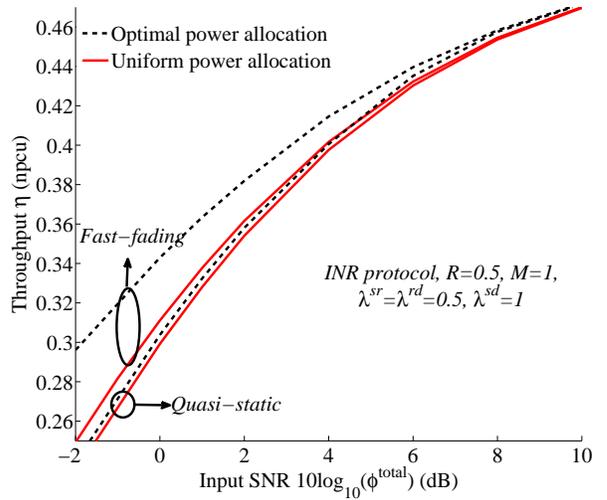}\\\vspace{-0mm}
\caption{The throughput of the relay-ARQ setup in different fading conditions, INR protocol, $R=0.5,\,M=1,$ $\lambda^\text{sd}=1,$ $\lambda^\text{sr}=\lambda^\text{rd}=0.5.$}\label{figure111}
\vspace{-0mm}
\end{figure}
Finally, it is interesting to note that, in harmony with \cite{throughputdef,6566132,Tcomkhodemun}, in all cases better system performance is achieved by the INR ARQ protocol, compared to the RTD.

\vspace{-0mm}
\section{Performance analysis in spatially-correlated fading conditions}
This section studies the system performance in spatially-correlated conditions.
As demonstrated, the discussions of the previous sections are helpful for performance analysis in spatially-correlated relay-ARQ setups.

For simplicity, we concentrate on the scenario where the relay is close to the destination, while the source is far from them. As an example of such a case, we can consider the moving-relay systems, e.g., \cite{6240247,6379135}. Also, the same discussions are valid for the scenario with the source and the relay close to each other.


Assuming the relay to be close to the destination, the relation between two fading random variables $h^\text{sd}$ and $h^\text{sr}$ is modeled by
\begin{align}\label{eq:correlatedmodel}
h^\text{sd}=\delta h^\text{sr}+\sqrt{1-\delta^2}\varsigma,\,\varsigma\sim\mathcal{CN}(0,\frac{1}{\lambda}).
\end{align}
Here, $\varsigma$ is a complex Gaussian variable $\mathcal{CN}(0,\frac{1}{\lambda})$ uncorrelated with $h^\text{sr}.$ Also, $\delta$ is a known correlation factor which demonstrates the dependency of the fading realizations. Moreover, $\lambda$ is the fading parameter of the source-relay and the source-destination links. This is a well-established model for the correlated Rayleigh-fading channels \cite{cog2,cog3,cog4}. Under this model, the joint and marginal pdfs are found as
\begin{align}\label{eq:jointpdf}
f_{g^\text{sd},g^\text{sr}}(x,y)=\frac{\lambda^2}{1-\delta^2}e^{-\lambda\frac{x+y}{1-\delta^2}}\mathcal{B}_0(\frac{2\lambda\delta\sqrt{xy}}{1-\delta^2}),
\end{align}
\begin{align}
f_{g^\text{sr}}(x)=\lambda e^{-\lambda x},\, f_{g^\text{sd}}(x)=\lambda e^{-\lambda x},
\end{align}
respectively, where $\mathcal{B}_0$ is the zeroth-order modified Bessel function of the first kind \cite{846501}. Also, we have $f_{g^\text{rd}}(x)=\lambda^\text{rd}e^{-\lambda^\text{rd}x}$, which is independent of the two other variables.

To study the effect of the spatial correlation, the key points are that 1) the two first steps of finding the closed-form expressions (\ref{eq:throughputdef})-(\ref{eq:powertotal}) are independent of the fading model. 2) The only terms which are affected by the fading model are the probabilities $\Pr(A_m),\Pr(B_{n,m}), \Pr(S_m).$ Moreover, these probabilities can be represented as $\pi=\Pr(g^\text{sr}\in [b^{\text{sr},1},b^{\text{sr},2}]\,\cap\, g^\text{sd}\in [b^{\text{sd},1},b^{\text{sd},2}]\,\cap \,g^\text{rd}\in [b^{\text{rd},1},b^{\text{rd},2}])$ with proper selection of the boundaries $b^{\text{sr},1},b^{\text{sr},2},b^{\text{sd},1},b^{\text{sd},2},b^{\text{rd},1},b^{\text{rd},2}.$ Thus, to take the spatial correlation into account, it is only required to calculate $\pi$ as
\begin{align}\label{eq:calpi}
&\pi=\int_{x=b^{\text{rd},1}}^{x=b^{\text{rd},2}}{f_{g^\text{rd}}(x)y(x)\text{d}x},
\nonumber\\&y(x)=\Pr\left(g^\text{sr}\in [b^{\text{sr},1},b^{\text{sr},2}]\cap g^\text{sd}\in [b^{\text{sd},1},b^{\text{sd},2}]\bigg|g^\text{rd}=x\right),
\end{align}
where $y(x)$ is determined based on the following procedure
\vspace{-0mm}
\begin{align}\label{eq:pdftwodimsubsampling}
\begin{array}{l}
 \Pr \{ {g^\text{sd}} \in [u,v)\,\cap {g^\text{sr}} \in [w,z)\,\} =\int_u^v {\int_w^z {{f_{{g^\text{sd}},{g^\text{sr}}}}(x,y)\text{d}x\text{d}y} }\\ \mathop  = \limits^{(a)} \int_u^v {\lambda{e^{ - \frac{x}{q}}}\bigg(\int_{\sqrt {\frac{{2w}}{q}} }^{\sqrt {\frac{{2z}}{q}} } {\varrho {e^{ - \frac{{{\varrho ^2}}}{2}}}{\mathcal{B} _0}(\chi{\sqrt {x} }\varrho )} \text{d}\varrho \bigg)\text{d}x}
 \\\mathop  = \limits^{(b)} \int_u^v {\lambda{e^{ - {\lambda x}}}\{ \mathcal{M} (\chi\sqrt x ,\sqrt {\frac{{2w}}{q}} ) - \mathcal{M} (\chi\sqrt x ,\sqrt {\frac{{2z}}{q}} )\} \text{d}x}\\
 \mathop  = \limits^{(c)} (1 - {\delta ^2}){e^{ - {\lambda w}}}\{ \mathcal{M} (\sqrt {\frac{{2w}}{q}} \delta ,\sqrt {\frac{{2u}}{q}} ) - \mathcal{M} (\sqrt {\frac{{2w}}{q}} \delta ,\sqrt {\frac{{2v}}{q}} )\}
  \\- (1 - {\delta ^2}){e^{ - {\lambda z}}}\{ \mathcal{M} (\sqrt {\frac{{2z}}{q}} \delta ,\sqrt {\frac{{2u}}{q}} ) - \mathcal{M} (\sqrt {\frac{{2z}}{q}} \delta ,\sqrt {\frac{{2v}}{q}} )\}\\
  + \lambda\int_u^v {{e^{ - {\lambda x}}}\{ \mathcal{M} (\sqrt {\frac{{2z}}{q}} ,\chi\sqrt x ) - \mathcal{M} (\sqrt {\frac{{2w}}{q}} ,\chi\sqrt x )\} } \text{d}x \\
 \mathop  = \limits^{(d)} {e^{ - {\lambda w}}}\{ \mathcal{M} (\sqrt {\frac{{2w}}{q}} \delta ,\sqrt {\frac{{2u}}{q}} ) - \mathcal{M} (\sqrt {\frac{{2w}}{q}} \delta ,\sqrt {\frac{{2v}}{q}} )\}\\
  - {e^{ - {\lambda z}}}\{ \mathcal{M} (\sqrt {\frac{{2z}}{q}} \delta ,\sqrt {\frac{{2u}}{q}} ) - \mathcal{M} (\sqrt {\frac{{2z}}{q}} \delta ,\sqrt {\frac{{2v}}{q}} )\}  \\
  + {e^{ - {\lambda v}}}\mathcal{M} (\sqrt {\frac{{2w}}{q}} ,\sqrt {\frac{{2v}}{q}} \delta ) - {e^{ - {\lambda u}}}\mathcal{M} (\sqrt {\frac{{2w}}{q}} ,\sqrt {\frac{{2u}}{q}} \delta )
  \\- {e^{ - {\lambda v}}}\mathcal{M} (\sqrt {\frac{{2z}}{q}} ,\sqrt {\frac{{2v}}{q}} \delta ) + {e^{ - {\lambda u}}}\mathcal{M} (\sqrt {\frac{{2z}}{q}} ,\sqrt {\frac{{2u}}{q}} \delta )\\=Y(u,v,w,z). \\
 \end{array}
\end{align}
Here, $(a)$ is obtained by defining $q \buildrel\textstyle.\over= \frac{1 - {\delta ^2}}{\lambda}$, $\chi \buildrel\textstyle.\over= \sqrt {{2 \mathord{\left/
 {\vphantom {2 r}} \right.
 \kern-\nulldelimiterspace} q}} \delta$ and using variable transform $\varrho  = \sqrt {{{2y} \mathord{\left/
 {\vphantom {{2y} r}} \right.
 \kern-\nulldelimiterspace} q}}
$. Then, $(b)$ is directly obtained from the definition of the Marcum Q-function $\mathcal{M} (x,y) = \int_y^\infty  {t{e^{ - \frac{{{t^2} + {x^2}}}{2}}}{\mathcal{B} _0}(xt)\text{d}t}.$
Finally, $(c)$ is based on the fact that $\mathcal{M} (x,y) = 1 + {e^{ - {{({x^2} + {y^2})} \mathord{\left/
 {\vphantom {{({x^2} + {y^2})} 2}} \right.
 \kern-\nulldelimiterspace} 2}}}{\mathcal{B} _0}(xy) - \mathcal{M} (y,x)$ and $(d)$ is derived by using variable transform $t = \sqrt x$, partial integration and some calculations.

As an example of (\ref{eq:calpi}), consider the relay-RTD scheme in spatially-correlated quasi-static channel condition where, using (\ref{eq:RTDSm}), (\ref{eq:RTDprobAm}), (\ref{eq:pdftwodimsubsampling}), the terms $\alpha_m,\beta_m,\gamma_M, \varepsilon_{j,m}$ are rephrased as
\begin{align}\label{eq:correlalpha}
\alpha_m&=\Pr(g^\text{sr}\in[\frac{e^R-1}{\sum_{i=1}^{m}{P_i^\text{s}}},\frac{e^R-1}{\sum_{i=1}^{m-1}{P_i^\text{s}}}]\cap g^\text{sd}\in[0,\frac{e^R-1}{\sum_{i=1}^{m}{P_i^\text{s}}}])\nonumber\\&=Y(0,\frac{e^R-1}{\sum_{i=1}^{m}{P_i^\text{s}}},\frac{e^R-1}{\sum_{i=1}^{m}{P_i^\text{s}}},\frac{e^R-1}{\sum_{i=1}^{m-1}{P_i^\text{s}}}),
\end{align}
\begin{align}\label{eq:correlbeta}
\beta_m=Y(\frac{e^R-1}{\sum_{i=1}^{m}{P_i^\text{s}}},\frac{e^R-1}{\sum_{i=1}^{m-1}{P_i^\text{s}}},0,\frac{e^R-1}{\sum_{i=1}^{m-1}{P_i^\text{s}}}),
\end{align}
\begin{align}\label{eq:correlgamma}
\gamma_M=Y(0,\frac{e^R-1}{\sum_{i=1}^{M}{P_i^\text{s}}} ,0,\frac{e^R-1}{\sum_{i=1}^{M}{P_i^\text{s}}}),
\end{align}
\begin{align}\label{eq:correlvarepsilon}
&\varepsilon_{j,m}=\int_{x=\frac{e^R-1}{\sum_{i=j+1}^{m}{P_i^\text{s}}}}^{\frac{e^R-1}{\sum_{i=j+1}^{m-1}{P_i^\text{s}}}}{\lambda^\text{rd}e^{-\lambda^\text{rd} x}Y\Bigg(\frac{e^R-1-x\sum_{i=j+1}^m{P_i^\text{r}}}{\sum_{i=1}^{j}{P_i^\text{s}}},}\nonumber\\&
\,\,\,\,\,\,\,\,\,\,\,\,\,\,\,\,\,\,\,\,\,\,\,\frac{e^R-1-x\sum_{i=j+1}^{m-1}{P_i^\text{r}}}{\sum_{i=1}^{j}{P_i^\text{s}}},\frac{e^R-1}{\sum_{i=1}^{j}{P_i^\text{s}}},\frac{e^R-1}{\sum_{i=1}^{j-1}{P_i^\text{s}}}\Bigg)\text{d}x.
\end{align}
Here, (\ref{eq:correlvarepsilon}) is obtained numerically with $Y(u,v,w,z)$ given in (\ref{eq:pdftwodimsubsampling}). Note that (\ref{eq:correlvarepsilon}) is one-dimensional integration with known integration boundaries. Thus, $\varepsilon_{j,m}$ is calculated easily. Also, the other probability terms, e.g., $\vartheta_{n}$, are obtained with the same procedure as in (\ref{eq:correlvarepsilon}). Then, having the probabilities, the system performance is studied with the same procedure as before.

Considering (\ref{eq:correlatedmodel}), Figs. 14-15 show the relative coverage region and the outage probability of the relay-ARQ setup for different correlation conditions. The results indicate that in the practical range of correlation conditions the fading dependencies do not affect the system performance considerably, in the sense that the coverage region and the outage probability changes are negligible at low $\delta$'s (The same point is valid for the throughput although not demonstrated in the figures). On the other hand, the data transmission efficiency of the relay-assisted network is considerably reduced at highly-correlated conditions, i.e., with $\delta\sim 1.$ Specifically, the relay network is mapped to the source-destination single-user setup if $\delta=1$ (please see (\ref{eq:correlatedmodel})). The coverage region increases with the number of retransmissions substantially (Fig. 14). Moreover, although not demonstrated in the figure, the results indicate that the sensitivity of the coverage region to the fading parameter $\lambda^\text{rd}$ (resp. correlation factor $\delta$) decreases with $\delta$ (resp. input SNR).

Finally, note that the performance gain of the relay-ARQ is at the cost of coordination overhead between the source and the relay. Particularly, the source data retransmission is decided based on the feedback signals from the destination and the relay, as opposed to non-relay setups with feedback only from the destination. Moreover, in harmony with every cooperative system, the mismatches between the source and the relay and the feedback delay may affect the data transmission efficiency of the relay-ARQ protocols. However, as shown in Fig. 11, the system performance is (almost) insensitive to the imperfect feedback signals for the practical range of the feedback error probabilities.


\begin{figure}\label{fig:NPA}
\vspace{-0mm}
\centering
  \includegraphics[width=.98\columnwidth]{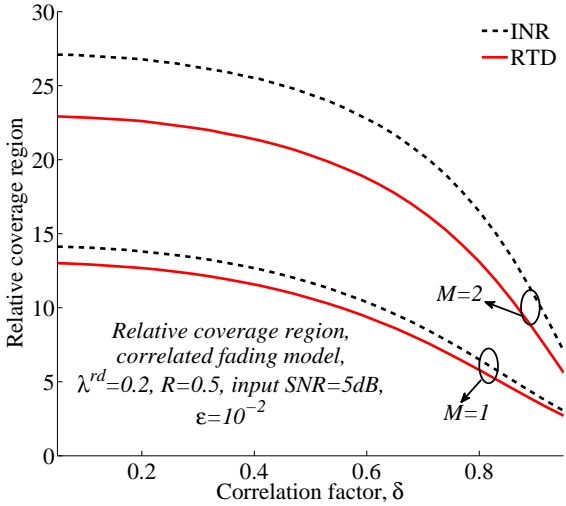}\\\vspace{-0mm}
\caption{Relative coverage region, compared to the single-user setup without ARQ, vs the correlation factor $\delta$ in (\ref{eq:correlatedmodel}). Input SNR=5dB, $R=0.5$, $\epsilon=10^{-2}$, $\lambda^\text{rd}=0.2.$}\label{figure111}
\vspace{-0mm}
\end{figure}

\begin{figure}\label{fig:NPA}
\vspace{-0mm}
\centering
  \includegraphics[width=.98\columnwidth]{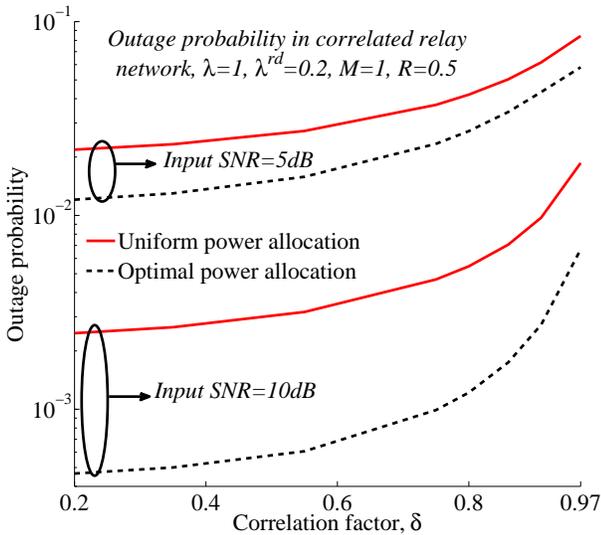}\\\vspace{-0mm}
\caption{Outage probability vs the correlation factor $\delta$ in (\ref{eq:correlatedmodel}). $\lambda=1,$ $\lambda^\text{rd}=0.2.$ RTD scheme with $R=0.5$, $M=1$. }\label{figure111}
\vspace{-0mm}
\end{figure}

\vspace{-3mm}
\section{Conclusion}
In this paper, we studied the performance of the relay-ARQ networks using adaptive power allocation.
The throughput and the outage probability were analyzed with different sum and individual power constraints on the source and the relay,  under independent and spatially-correlated fading conditions. The results show that considerable outage probability reduction and throughput/coverage region increment are achieved by optimal power allocation between the source and the relay. The performance of the relay-ARQ network is not sensitive to spatial correlation, within the practical range. Also, the effect of imperfect feedback signals on the data transmission efficiency of the relay-ARQ scheme is negligible, if the feedback bit error probability is in the practical range of interest. Finally, substantial performance improvement is achieved by variable-length coding and increasing the number of ARQ-based retransmissions.
\vspace{-3mm}
\section{Appendix}
\subsection{Deriving the probability terms for the INR protocol in quasi-static channels}
Considering the INR, the probabilities $\Pr(A_m),\,\Pr(B_{n,m}),\,\Pr(S_m)$ are determined with the same procedure as for the RTD protocol, i.e., (\ref{eq:RTDSm})-(\ref{eq:RTDprobBnm}), while the terms $\alpha_m,\,\beta_m,\,\gamma_M,\,\omega_j,\,\theta_{j,m}$ and $\rho_n$ are recalculated based on the maximum decodable rate functions of the INR protocol. For instance, using (\ref{eq:functionINRR})-(\ref{eq:functionINRD}), the probability $\alpha_m$ in (\ref{eq:RTDSm}) is rephrased as
\vspace{-2mm}
\begin{align}\label{eq:alphaINR}
&\alpha_m=\Pr(\sum_{i=1}^{m-1}{\frac{{l_i\log(1+g^\text{sr}P_i^\text{s})}}{\sum_{j=1}^{m-1}{l_j}}}<R_{(m-1)}\,\nonumber\\&\cap  \sum_{i=1}^{m}{\frac{{l_i\log(1+g^\text{sr}P_i^\text{s})}}{\sum_{j=1}^{m}{l_j}}}\ge R_{(m)}
\,\cap  \sum_{i=1}^{m}{\frac{{l_i\log(1+g^\text{sd}P_i^\text{s})}}{\sum_{j=1}^{m}{l_j}}}<R_{(m)}  )
\nonumber\\&
=\Pr(\forall g^\text{sr},g^\text{sd}|   U_{(m-1)}^\text{r}(g^\text{sr})<R_{(m-1)} \,\nonumber\\&\,\,\,\,\,\,\,\,\,\,\,\,\,\,\,\,\cap U_{(m)}^\text{r}(g^\text{sr})\ge R_{(m)} \,\cap   U_{(m,m)}^\text{d}(g^\text{sd})< R_{(m)}    )\nonumber\\&=(e^{-\lambda^\text{sr}x_{(m)}^\text{r}}-e^{-\lambda^\text{sr}x_{(m-1)}^\text{r}})(1-e^{-\lambda^\text{sd}x_{(m,m)}^\text{d}})
\end{align}
for the INR. Here, we have defined
\vspace{-0mm}
\begin{align}\label{eq:defsolution1}
x_{(m)}^\text{r}\doteq \mathop {\arg  }\limits_{x} \,\{ U_{(m)}^\text{r}(x)= R_{(m)} \},
\end{align}
\vspace{-0mm}
\begin{align}\label{eq:defsolution2}
x_{(m,m)}^\text{d}\doteq \mathop {\arg  }\limits_{x} \,\{ U_{(m,m)}^\text{d}(x)= R_{(m)} \}.
\end{align}
Also, with the same arguments, the terms $\beta_m,$ $\gamma_M$ and $\omega_j$ in (\ref{eq:RTDSm}) and (\ref{eq:varepsiloncal}) are obtained by
\vspace{-0mm}
\begin{align}\label{eq:betaINR}
\beta_m=(e^{-\lambda^\text{sd}x_{(m,m)}^\text{d}}-e^{-\lambda^\text{sd}x_{(m-1,m-1)}^\text{d}})(1-e^{-\lambda^\text{sr}x_{(m-1)}^\text{r}}),
\end{align}
\vspace{-0mm}
\begin{align}
\gamma_M=(1-e^{-\lambda^\text{sd}x_{(M,M)}^\text{d}})(1-e^{-\lambda^\text{sr}x_{(M)}^\text{r}}),
\end{align}
\vspace{-0mm}
\begin{align}
\omega_j=e^{-\lambda^\text{sr}x_{(j)}^\text{r}}-e^{-\lambda^\text{sr}x_{(j-1)}^\text{r}},
\end{align}
when the INR ARQ is implemented. Note that $U_{(m)}^\text{r}(0)=0$ and $U_{(m,m)}^\text{d}(0)=0$. Moreover, as $R_{(i)}<R_{(i-1)},\forall i,$ it is straightforward to show that $U_{(m)}^\text{r}(x)$ and $U_{(m,m)}^\text{d}(x)$, i.e., (\ref{eq:functionINRR}) and (\ref{eq:functionINRD2}), are increasing functions of $x$. Therefore, the solutions of (\ref{eq:defsolution1})-(\ref{eq:defsolution2}), i.e., $x_{(m)}^\text{r}$ and $x_{(m,m)}^\text{d},$ will be unique for any given values of $R_{(m)},P_i^\text{s},i=1,\ldots,m.$ Unfortunately, to the best of authors' knowledge, there is no closed-form solution for $x_{(m)}^\text{r}$ and $x_{(m,m)}^\text{d}$. However, because of the uniqueness property, the solutions of (\ref{eq:defsolution1})-(\ref{eq:defsolution2}) are easily obtained via numerical methods, such as the ``fsolve'' function of MATLAB.

Finally, using the INR, the probabilities $\theta_{j,m}$ and $\rho_n$ should be calculated based on, e.g.,
\begin{align}
\rho_n=\lambda^\text{sd}\lambda^\text{rd}\int\int_{U_{n,M}^\text{d}(x,y)\le R_{(M)}}{e^{-(\lambda^\text{sd}x+\lambda^\text{rd}y)}\text{d}x\text{d}y},
\end{align}
which is a two-dimensional numerical integration and, consequently, difficult to find. This is because the boundaries of the two-dimensional integrals can not be expressed in closed-form. To tackle the problem, two different methods can be used. The first one is the bounds introduced in Theorems 1-2. The second method is to use $\log(1+x)\simeq x$ for low SNRs which leads to the following approximation
\vspace{-0mm}
\begin{align}
\rho_n&=\Pr\Bigg(\sum_{i=1}^n{\left((\frac{1}{R_{(i)}}-\frac{1}{R_{(i-1)}})\log(1+P_i^\text{s}g^\text{sd})\right)}\nonumber\\&\,\,\,\,\,\,\,\,\,\,\,\,\,\,+\sum_{i=n+1}^M{\left((\frac{1}{R_{(i)}}-\frac{1}{R_{(i-1)}})\log(1+g^\text{rd}P_i^\text{s})\right)}\le 1\Bigg)\nonumber\\&\simeq\Pr\Bigg(g^\text{sd}\sum_{i=1}^n{\left((\frac{1}{R_{(i)}}-\frac{1}{R_{(i-1)}})P_i^\text{s}\right)}\nonumber\\&\,\,\,\,\,\,\,\,\,\,\,\,\,\,\,\,+g^\text{rd}\sum_{i=n+1}^M{\left((\frac{1}{R_{(i)}}-\frac{1}{R_{(i-1)}})P_i^\text{s}\right)}\le 1\Bigg)\nonumber\\&=\int_0^\frac{1}{\sum_{i=1}^n{((\frac{1}{R_{(i)}}-\frac{1}{R_{(i-1)}})P_i^\text{s})}}{f_{g^\text{sd}}(x)\times}\nonumber\\&\,\,\,\,\,\,\,\,\,\,\,\,\,\,\,\,\,\,\,{\Pr\left(g^\text{rd}\le \frac{1-\sum_{i=1}^n{((\frac{1}{R_{(i)}}-\frac{1}{R_{(i-1)}})P_i^\text{s})}x}{\sum_{i=n+1}^M{(\frac{1}{R_{(i)}}-\frac{1}{R_{(i-1)}})P_i^\text{s}}}\right)\text{d}x}
\nonumber\\&
=\lambda^\text{sd}\int_0^\frac{1}{\sum_{i=1}^n{(\frac{1}{R_{(i)}}-\frac{1}{R_{(i-1)}})P_i^\text{s}}}{e^{-\lambda^\text{sd}x}\times}\nonumber\\&\,\,\,\,\,\,\,\,\,\,\,\,\,\,\,\,\,\,\,\,\,\,\,\,\,\,\,\,\,\,\,\,\,\,\,\,\,{\left(1-e^{-\lambda^\text{rd}(\frac{1-{\sum_{i=1}^n{((\frac{1}{R_{(i)}}-\frac{1}{R_{(i-1)}})P_i^\text{s})}}x}{\sum_{i=n+1}^M{(\frac{1}{R_{(i)}}-\frac{1}{R_{(i-1)}})P_i^\text{s}}})}\right)\text{d}x}
\nonumber\\&
=1-\frac{e^{-\frac{\lambda^\text{rd}}{\sum_{i=n+1}^M{((\frac{1}{R_{(i)}}-\frac{1}{R_{(i-1)}})P_i^\text{s})}}}}{1-\frac{\lambda^\text{rd}{\sum_{i=1}^n{((\frac{1}{R_{(i)}}-\frac{1}{R_{(i-1)}})P_i^\text{s})}}}{\lambda^\text{sd}{\sum_{i=n+1}^M{((\frac{1}{R_{(i)}}-\frac{1}{R_{(i-1)}})P_i^\text{s})}}}}\nonumber\\&-\left(1-\frac{1}{1-\frac{\lambda^\text{rd}{\sum_{i=1}^n{((\frac{1}{R_{(i)}}-\frac{1}{R_{(i-1)}})P_i^\text{s})}}}{\lambda^\text{sd}{\sum_{i=n+1}^M{((\frac{1}{R_{(i)}}-\frac{1}{R_{(i-1)}})P_i^\text{s})}}}}\right)\times \nonumber\\&e^{-\frac{\lambda^\text{sd}}{\sum_{i=1}^n{((\frac{1}{R_{(i)}}-\frac{1}{R_{(i-1)}})P_i^\text{s})}}}.
\end{align}
The same approach can be used to find $\theta_{j,m}$ for low SNRs. Finally, having $\Pr(A_m),\,\Pr(B_{n,m}),\,\Pr(S_m)$ as functions of $R_{(m)},\,P_m^\text{s},\,P_m^\text{r},m=1,\ldots,M+1,$ the system performance can be studied with the same procedure as before (please see subsection IV.D).
\vspace{-0mm}
\subsection{Proof of Theorem 1}

Considering (\ref{eq:RTDSm})-(\ref{eq:UUfixedlength}), it can be easily shown that the performance of the fixed-length INR scheme is a decreasing function of the probability terms $\Pr(\sum_{i=1}^m{\log(1+g^\text{sr}P_i^\text{s})}\le R)$ and $\Pr(\sum_{i=1}^j{\log(1+g^\text{sd}P_i^\text{s})}+\sum_{i=j+1}^m{\log(1+g^\text{rd}P_i^\text{r})}\le R)$. In other words, the system performance is underestimated if the maximum decodable rates $U_m^\text{r, fixed-length}$ and $U_{j,m}^\text{d, fixed-length}$ are replaced by their corresponding lower bounds. From (\ref{eq:Ufixedlength}), we can write
\vspace{-0mm}
\begin{align}\label{eq:theorem1}
\Pr\left(U_m^\text{r, fixed-length}(g^\text{sr})\le \frac{R}{m}\right)&=\Pr\left(\sum_{i=1}^m{\log(1+g^\text{sr}P_i^\text{s})}\le R\right)\nonumber\\&=\Pr(\Psi\le e^R).
\end{align}
Here, $\Psi$ is defined as
\vspace{-0mm}
\begin{align}\label{eq:defPsi}
\Psi \doteq\prod_{i=1}^{m}{\left(1+g^\text{sr}P_i^\text{s}\right)}=\det(\textbf{I}_m+\textbf{C})
\end{align}
with $\textbf{I}_m$ representing the $m\times m$ identity matrix and $\textbf{C}=[c_{i,k}]$ denoting the diagonal matrix given by\footnote{The matrices are presented by capital bold letters.}
\vspace{-0mm}
\begin{align}\label{eq:defmatC}
&c_{i,k} = \left\{ \begin{array}{l}
 {g^\text{sr}P_i^\text{s}}\,\,\,\,\,\,\,\,\text{if}\,\,i=k,i=1,\ldots,m, \\
 0\,\,\,\,\,\,\,\,\,\,\,\,\,\,\,\,\,\,\text{if}\,\,i\ne k. \\
 \end{array} \right.
\end{align}
Using the Minkowski's inequality \cite[Theorem 7.8.8]{minkowskibook} in (\ref{eq:defPsi}) leads to
\vspace{-0mm}
\begin{align}
\Psi =\det(\textbf{I}_m+\textbf{C})\ge (1+\det(\textbf{C})^\frac{1}{m})^m.
\end{align}
Thus, from $\det(\textbf{C})=(g^\text{sr})^m\prod_{i=1}^m{P_i^\text{s}}$, we have $\Psi\ge (1+g^\text{sr}\sqrt[m]{\prod_{i=1}^m{P_i^\text{s}}})^m$ and
\vspace{-0mm}
\begin{align}\label{eq:Probapprox1}
&\Pr\left(\sum_{i=1}^m{\log(1+g^\text{sr}P_i^\text{s})}\le R\right)\le\nonumber\\& \Pr\left(\left(1+g^\text{sr}\sqrt[m]{\prod_{i=1}^m{P_i^\text{s}}}\right)^m \le e^R\right)=F_{g^\text{sr}}(\frac{e^{\frac{R}{m}}-1}{\sqrt[m]{\prod_{i=1}^m{P_i^\text{s}}}}),
\end{align}
as stated in the theorem.

For the second inequality of the theorem, i.e., (\ref{eq:theoremstate2}), the same arguments as in (\ref{eq:theorem1})-(\ref{eq:Probapprox1}) are used to write
\vspace{-0mm}
\begin{align}
&\Pr\left(\sum_{i=1}^j{\log(1+g^\text{sd}P_i^\text{s})}+\sum_{i=j+1}^m{\log(1+g^\text{rd}P_i^\text{r})}\le R\right)\nonumber\\&=\Pr(\Upsilon\le e^R),
\nonumber\\&
\Upsilon= \prod_{i=1}^j{(1+g^\text{sd}P_i^\text{s})}\prod_{i=j+1}^m{(1+g^\text{rd}P_i^\text{r})}=\det(\textbf{I}_m+\textbf{D}),
\end{align}
where $\textbf{D}=[d_{k,n}]$ is the $m\times m$ diagonal matrix defined by
\vspace{-0mm}
\begin{align}
d_{k,n} = \left\{ \begin{array}{l}
 {g^\text{sd}P_i^\text{s}}\,\,\,\,\,\,\,\,\text{if}\,\,k=n,k=1,\ldots,j, \\
{g^\text{rd}P_i^\text{r}}\,\,\,\,\,\,\,\,\text{if}\,\,k=n,k=j+1,\ldots,m, \\
 0\,\,\,\,\,\,\,\,\,\,\,\,\,\,\,\,\,\,\text{if}\,\,k\ne n.\nonumber \\
 \end{array} \right.
\end{align}
In this way, we reuse the Minkowski's inequality to write
\vspace{-0mm}
\begin{align}
\Upsilon&\ge (1+\det(\textbf{D})^\frac{1}{m})^m\nonumber\\&=\left(1+(g^\text{sd})^{\frac{j}{m}}(g^\text{rd})^{\frac{m-j}{m}}\sqrt[m]{\prod_{i=1}^j{P_i^\text{s}}\prod_{i=j+1}^m{P_i^\text{r}}}\right)^m
\end{align}
which, from the definition of $V_{j,m}(v)$ in (\ref{eq:theoremstate2}), leads to
\vspace{-0mm}
\begin{align}
\begin{array}{l}
\Pr(\Upsilon\le e^R)\le\Pr\left((g^\text{sd})^{\frac{j}{m}}(g^\text{rd})^{\frac{m-j}{m}}\le \frac{e^\frac{R}{m}-1}{\sqrt[m]{\prod_{i=1}^j{P_i^\text{s}}\prod_{i=j+1}^m{P_i^\text{r}}}}\right)\\\,\,\,\,\,\,\,\,\,\,\,\,\,\,\,\,\,\,\,\,\,\,\,\,\,\,\,\,\,\,\,\,\,=\int_0^\infty{f_{g^\text{sd}}(x)\Pr(g^\text{rd}\le \frac{s}{x^\frac{j}{m-j}})\text{d}x}\\\,\,\,\,\,\,\,\,\,\,\,\,\,\,\,\,\,\,\,\,\,\,\,\,\,\,\,\,\,\,\,\,\,=1-\lambda^\text{sd}\int_0^\infty{e^{-\lambda^\text{sd}x-\lambda^\text{rd}x^\frac{j}{j-m}s}\text{d}x}\\\,\,\,\,\,\,\,\,\,\,\,\,\,\,\,\,\,\,\,\,\,\,\,\,\,\,\,\,\,\,\,\,=1-V_{j,m}(s),s=\left(\frac{e^{\frac{R}{m}}-1}{\sqrt[m]{\prod_{i=1}^j{P_i^\text{s}}\prod_{i=j+1}^m{P_i^\text{r}}}}\right)^\frac{m}{m-j}.\\
 \end{array}
\end{align}
\vspace{-0mm}
\subsection{Proof of Theorem 2}
To prove the theorem, the following modifications are applied into the arguments of Theorem 1. Considering (\ref{eq:defPsi}), we  rewrite $\Psi$ as
\vspace{-0mm}
\begin{align}\label{eq:proofth21}
\Psi=\prod_{i=1}^{m}{(1+g^\text{sr}P_i^\text{s})}=\det(\textbf{I}_m+\textbf{G}^\text{sr}\textbf{P}^\text{s}).
\end{align}
Here, $\textbf{G}^\text{sr}\doteq g^\text{sr}\textbf{I}_m$ and $\textbf{P}^\text{s}\doteq\frac{1}{g^\text{sr}}\textbf{C}$ where $\textbf{C}$ is given in (\ref{eq:defmatC}). By setting $\textbf{B}=\textbf{I}_m, \textbf{A}=\textbf{G}^\text{sr},$ $\textbf{X}=\textbf{P}^\text{s},$ denoting the conjugate transpose of the matrix $\textbf{X}$ by $\textbf{X}^*$ and because the matrices $\textbf{G}^\text{sr}$ and $\textbf{P}^\text{s}$ are Hermitian, we use
\begin{align}\label{eq:secondinequality}
\det(\textbf{AA}^*+\textbf{BB}^*)\det(\textbf{I}_m+\textbf{X}^*\textbf{X})\ge (\det(\textbf{B}+\textbf{AX}))^2
\end{align}
\cite[Theorem 3.4]{inequalitytheorem} to write
\begin{align}
\Psi&\le \sqrt{ \det((\textbf{G}^\text{sr})^2+\textbf{I}_m) \det(\textbf{I}_m+(\textbf{P}^\text{s})^2)    }\nonumber\\&=\sqrt{(1+(g^\text{sr})^2)^m\prod_{i=1}^m{(1+(P_i^\text{s})^2)}}.
\end{align}
Therefore, a lower bound of the probability $\Pr\left(\sum_{i=1}^m{\log(1+g^\text{sr}P_i^\text{s})}\le R\right)$ is obtained by
\begin{align}\label{eq:Probapprox21}
&\Pr\left(\sum_{i=1}^m{\log(1+g^\text{sr}P_i^\text{s})}\le R\right)\nonumber\\&\ge \Pr\left(\sqrt{(1+(g^\text{sr})^2)^m\prod_{i=1}^m{(1+(P_i^\text{s})^2)}} \le e^R\right)\nonumber\\&=F_{g^\text{sr}}(\sqrt{\frac{e^\frac{2R}{m}}{\sqrt[m]{\prod_{i=1}^m{(1+(P_i^\text{s})^2)}}}-1})\nonumber\\&=1-e^{-\lambda^\text{sr}\sqrt{\frac{e^\frac{2R}{m}}{\sqrt[m]{\prod_{i=1}^m{(1+(P_i^\text{s})^2)}}}-1}}.
\end{align}
For (\ref{eq:theoremstate22}), i.e., the second inequality of the theorem, we redefine $\textbf{B}=\textbf{I}_m,$   $\textbf{A}=\textbf{G}^\text{sd,rd}, \textbf{X}=\textbf{P}^\text{s,r}$ such that $\textbf{G}^\text{sd,rd}=[g_{k,n}^\text{sd,rd}]$  and $\textbf{P}^\text{s,r}=[p_{k,n}^\text{s,r}]$ with
\begin{align}
g_{k,n}^\text{sd,rd} = \left\{ \begin{array}{l}
 {g^\text{sd}}\,\,\,\,\,\,\,\,\text{if}\,\,k=n,k=1,\ldots,j, \\
{g^\text{rd}}\,\,\,\,\,\,\,\,\text{if}\,\,k=n,k=j+1,\ldots,m, \\
 0\,\,\,\,\,\,\,\,\,\,\,\,\text{if}\,\,k\ne n.\nonumber \\
 \end{array} \right.
\end{align}
\begin{align}
p_{k,n}^\text{s,r} = \left\{ \begin{array}{l}
 {P_i^\text{s}}\,\,\,\,\,\,\,\,\text{if}\,\,k=n,k=1,\ldots,j, \\
{P_i^\text{r}}\,\,\,\,\,\,\,\,\text{if}\,\,k=n,k=j+1,\ldots,m, \\
 0\,\,\,\,\,\,\,\,\,\,\,\,\text{if}\,\,k\ne n.\nonumber \\
 \end{array} \right.
\end{align}
Then, we reuse (\ref{eq:secondinequality}) to write
\begin{align}
\Upsilon &= \prod_{i=1}^j{(1+g^\text{sd}P_i^\text{s})}\prod_{i=j+1}^m{(1+g^\text{rd}P_i^\text{r})}=\det(\textbf{B}+\textbf{G}^\text{sd,rd}\textbf{P}^\text{s,r})\nonumber\\&\le \sqrt{\bigg(1+(g^\text{sd})^2\bigg)^j\bigg(1+(g^\text{rd})^2\bigg)^{m-j}\zeta},\nonumber\\&
\zeta\doteq\prod_{i=1}^j{\bigg(1+(P_i^\text{s})^2\bigg)}\prod_{i=j+1}^m{\bigg(1+(P_i^\text{r})^2\bigg)}.
\end{align}
Thus, the probability $\Pr(U_{j,m}^\text{d,fixed-length}\le \frac{R}{m})$ is lower bounded by
\begin{align}
&\Pr(\Upsilon\le e^R)\ge \Pr\bigg(\big(1+(g^\text{sd})^2\big)^j\big(1+(g^\text{rd})^2\big)^{m-j}\le r\bigg)\nonumber\\&=\int_0^{\sqrt{\sqrt[j]{r}-1}}{f_{g^\text{sd}}(x)\Pr\left(g^\text{rd}\le \sqrt{\sqrt[m-j]{r}(1+x^2)^\frac{j}{j-m}-1}\right)\text{d}x}\nonumber\\&=W_{j,m}(r), \nonumber\\& W_{j,m}(r)\doteq\int_0^{\sqrt{\sqrt[j]{r}-1}}{\lambda^\text{sd}e^{-\lambda^\text{sd}x}{\Big(1-e^{-\lambda^\text{rd}\sqrt{\sqrt[m-j]{r}(1+x^2)^\frac{j}{j-m}-1}}\Big)}\text{d}x},\nonumber\\&
r=\frac{e^{2R}}{\prod_{i=1}^j{(1+(P_i^\text{s})^2)}\prod_{i=j+1}^m{(1+(P_i^\text{r})^2)}},
\end{align}
if the transmission powers are so low (or the initial transmission rate $R$ is so high) that $r\ge 1.$
\vspace{-0mm}
\subsection{Proof of Theorem 3}
To obtain (\ref{eq:theoremeq31}), and the decoding probabilities of the RTD, let us define $Z_{j,m}=\sum_{i=1}^m{z_{j,m}(i)}$ with
\begin{align}\label{proofeq31}
z_{j,m}(i) = \left\{ \begin{array}{l}
 {g_i^\text{sd}P_i^\text{s}}\,\,\,\,\,\,\,\,\text{if}\,\,i=1,\ldots,j, \\
{g_i^\text{rd}P_i^\text{r}}\,\,\,\,\,\,\,\,\text{if}\,\,i=j+1,\ldots,m. \\
 \end{array} \right.
\end{align}
We have
\vspace{-0mm}
\begin{align}\label{proofeq32}
&\Pr(\log(1+\sum_{i=1}^j{P_i^\text{s}g_i^\text{sd}}+\sum_{i=j+1}^m{P_i^\text{r}g_i^\text{rd}})\le R)\nonumber\\&=\Pr(Z_{j,m}\le e^R-1)=\int_0^{e^R-1}{f_{Z_{j,m}}(x)\text{d}x},
\end{align}
where $f_{Z_{j,m}}$ is the pdf of the random variable ${Z_{j,m}}.$ As the pdf of the sum of independent random variables is obtained by the convolution of their pdfs, we use (\ref{proofeq31}) and the inverse Laplace transform $\mathcal{L}^{-1}\{.\}$ to write
\begin{align}\label{proofeq33}
f_{Z_{j,m}}(x)&\mathop  = \limits^{(a)}\mathcal{L}^{-1}\{\frac{1}{\prod_{i=1}^j{(1+\frac{P_i^\text{s}s}{\lambda^\text{sd}})}\prod_{i=j+1}^m{(1+\frac{P_i^\text{r}s}{\lambda^\text{rd}})}}\}\nonumber\\&\mathop  = \limits^{(b)}\mathcal{L}^{-1}\{\sum_{i=1}^j{\frac{a_i^\text{sd}}{1+\frac{P_i^\text{s}s}{\lambda^\text{sd}}}}+\sum_{i=j+1}^m{\frac{a_i^\text{rd}}{1+\frac{P_i^\text{r}s}{\lambda^\text{rd}}}}\}\nonumber\\&
=\sum_{i=1}^j{\frac{P_i^\text{s}a_i^\text{sd}}{\lambda^\text{sd}}e^{-\frac{\lambda^\text{sd}x}{P_i^\text{s}}}}+\sum_{i=j+1}^m{\frac{P_i^\text{r}a_i^\text{rd}}{\lambda^\text{rd}}e^{-\frac{\lambda^\text{rd}x}{P_i^\text{r}}}},\nonumber\\& a_i^\text{sd}={\frac{1}{\prod_{k=1,k\ne i}^j{(1-\frac{P_k^\text{s}}{P_i^\text{s}})}\prod_{k=j+1}^m{(1-\frac{P_k^\text{r}\lambda^\text{sd}}{\lambda^\text{rd}P_i^\text{s}})}}}, \nonumber\\&a_i^\text{rd}={\frac{1}{\prod_{k=1}^j{(1-\frac{P_k^\text{s}\lambda^\text{rd}}{\lambda^\text{sd}P_i^\text{r}})}\prod_{k=j+1,k\ne i}^m{(1-\frac{P_k^\text{r}}{P_i^\text{r}})}}}.
\end{align}
Here, $(a)$ is based on (\ref{proofeq31}) for independent Rayleigh-fading variables $g_i^\text{sd},\,g_i^\text{rd},i=1,\ldots,m,$ and
\begin{align}\label{Aproofeq31}
\mathcal{L}\{f_{z_{j,m}(i)}\} = \left\{ \begin{array}{l}
 \frac{1}{1+\frac{P_i^\text{s}s}{\lambda^\text{sd}}}\,\,\,\,\,\,\,\,\text{if}\,\,i=1,\ldots,j, \\
\frac{1}{1+\frac{P_i^\text{r}s}{\lambda^\text{rd}}}\,\,\,\,\,\,\,\,\text{if}\,\,i=j+1,\ldots,m. \\
 \end{array} \right.
\end{align}
Also, $(b)$ is obtained by partial fraction expansion of $\mathcal{D}(s)=\frac{1}{\prod_{i=1}^j{(1+\frac{P_i^\text{s}s}{\lambda^\text{sd}})}\prod_{i=j+1}^m{(1+\frac{P_i^\text{r}s}{\lambda^\text{rd}})}},\, P_i^\text{s}\ne P_k^\text{s}, P_i^\text{r}\ne P_k^\text{r},i\ne k,$ with $a_i^\text{sd}$ and $a_i^\text{rd},\,i=1,\ldots,m,$ representing the fraction expansion coefficients. Replacing (\ref{proofeq33}) into (\ref{proofeq32}) leads to (\ref{eq:theoremeq31}), as stated in the theorem. Note that (\ref{proofeq33}) is based on the assumption that the function $D(s)$ consists of $m$ first-order poles, which is the case with optimal power allocation. Straightforward modifications should be applied in (\ref{eq:theoremeq31}) and (\ref{proofeq33}), if $D(s)$ has poles of order $n>1$.

Equation (\ref{eq:theoremeq32}), on the other hand, is a direct consequence of \cite[Corollary 2]{5357980} which, due to space limit, is not repeated here. Finally, using the same procedure as in (\ref{eq:theoremeq31}) and (\ref{eq:theoremeq32}), we can find $\alpha_m,\,\beta_m,\,\gamma_M,\,\omega_j,\,\theta_{j,m},$ $\rho_n$ and the probability terms of the RTD and INR protocols, respectively.
\bibliographystyle{IEEEtran} 
\bibliography{masterRelayedited}

\vfill

\end{document}